\documentclass{nature-single}

\usepackage{lineno}

\usepackage[top=0.8in, bottom=1in, left=0.8in, right=0.8in]{geometry}

\usepackage{graphicx}
\usepackage{amsmath}
\usepackage{amssymb}
\usepackage{threeparttable}
\usepackage{caption}
\usepackage{bibunits}
\usepackage{color}
\usepackage{adjustbox}
\usepackage{url}
\usepackage{soul}
\usepackage[misc]{ifsym} 
\usepackage{bbding} 
\usepackage{threeparttablex, tablefootnote}
\usepackage{longtable}
\usepackage{booktabs}

\usepackage{graphics}
\usepackage{hyperref}
\usepackage[dvipsnames]{xcolor}
\usepackage{newtxtext}
\usepackage[varg,smallerops,upint]{newtxmath}
\usepackage{newfloat}
\DeclareFloatingEnvironment[fileext=lop]{Extended Data Figure}

\renewcommand{\arraystretch}{1.4} 
\defaultbibliography{MasterBiblio.bib}
\defaultbibliographystyle{naturemag}

\title{Black holes regulate cool gas accretion in massive galaxies}

\author{
Tao Wang$^{1,2}$, Ke Xu$^{1,2}$, Yuxuan Wu$^{1,2}$, Yong Shi$^{1,2}$, David Elbaz$^{3}$, Luis C. Ho$^{4,5}$, Zhi-Yu Zhang$^{1,2}$,  Qiusheng Gu$^{1,2}$, Yijun Wang$^{1,2}$, Chenggang Shu$^{6}$, Feng Yuan$^{7}$, Xiaoyang Xia$^{8}$ \& Kai Wang$^{4}$
\vspace{8pt}}

\begin{document}
\maketitle

\begin{affiliations}
\small
\item School of Astronomy and Space Science, Nanjing University, Nanjing, Jiangsu 210093, China
    \item Key Laboratory of Modern Astronomy and Astrophysics, Nanjing University, Ministry of Education, Nanjing 210093, China
    \item Universit\'e Paris-Saclay, Universit\'e Paris Cit\'e, CEA, CNRS, AIM, Gif-sur-Yvette, France
    \item Kavli Institute for Astronomy and Astrophysics, Peking University, Beijing 100871, China 
    \item Department of Astronomy, School of Physics, Peking University, Beijing 100871, China
    \item Shanghai Key Lab for Astrophysics, Shanghai Normal University,  Shanghai 200234,  China
    \item Center for Astronomy and Astrophysics and Department of Physics, Fudan University, Shanghai 200438, China
    \item Tianjin Astrophysics Center, Tianjin Normal University, Tianjin 300387, China
\Envelope e-mail: taowang@nju.edu.cn
\end{affiliations}

\begin{bibunit}

\begin{abstract}
The nucleus of almost all massive galaxies contains a supermassive black hole (BH)\cite{Ho:2008}. The feedback from the accretion of these BHs is often considered to have crucial roles in establishing the quiescence of massive galaxies\cite{Silk:1998,Balogh:2001,DiMatteo:2005,Cattaneo:2009,Fabian:2012,Maiolino:2012,Cicone:2014,Bluck:2014,Terrazas:2016b,Terrazas:2017,Piotrowska:2022,Bluck:2022,Brownson:2022}, although some recent studies show that even galaxies hosting the most active BHs do not exhibit a reduction in their molecular gas reservoirs or star formation rates\cite{Stanley:2017,Schulze:2019,Shangguan:2020}. Therefore, the influence of BHs on galaxy star formation remains highly debated and lacks direct evidence. Here, based on a large sample of nearby galaxies with measurements of masses of both BHs and atomic hydrogen (HI), the main component of the interstellar medium\cite{Saintonge:2017}, we show that the HI gas mass to stellar masses ratio ($\mu_{\rm HI} = M_{\rm HI}/M_{\star}$) is more strongly correlated with BH masses ($M_{\rm BH}$) than with any other galaxy parameters, including stellar mass, stellar mass surface density and bulge masses. Moreover, once the $\mu_{\rm HI}-M_{\rm BH}$ correlation is considered, $\mu_{\rm HI}$ loses dependence on other galactic parameters, demonstrating that $M_{\rm BH}$ serves as the primary driver of $\mu_{\rm HI}$. These findings provide important evidence for how the accumulated energy from BH accretion regulates the cool gas content in galaxies, by ejecting interstellar medium gas and/or suppressing gas cooling from the circumgalactic medium.
\end{abstract}

Our primary sample consists of 69 central galaxies in the nearby Universe with direct estimates of black hole (BH) masses derived from resolved kinematics of stars or gas\cite{Kormendy:2013,Saglia:2016,vandenBosch:2016,Terrazas:2017}. We have included only central galaxies to avoid any environmental impact on the interstellar medium (ISM) properties of galaxies. The sample includes several types of galaxy, including spirals, lenticulars and ellipticals. We obtained the atomic hydrogen (HI) 21-cm emission fluxes, which trace the atomic gas mass $M_{\rm HI}$, by crossmatching with nearby galaxy databases (Methods and Extended Data Table ~\ref{etb1}).

\begin{figure*}[h!]
	\centering
	\includegraphics[width=0.8\textwidth]{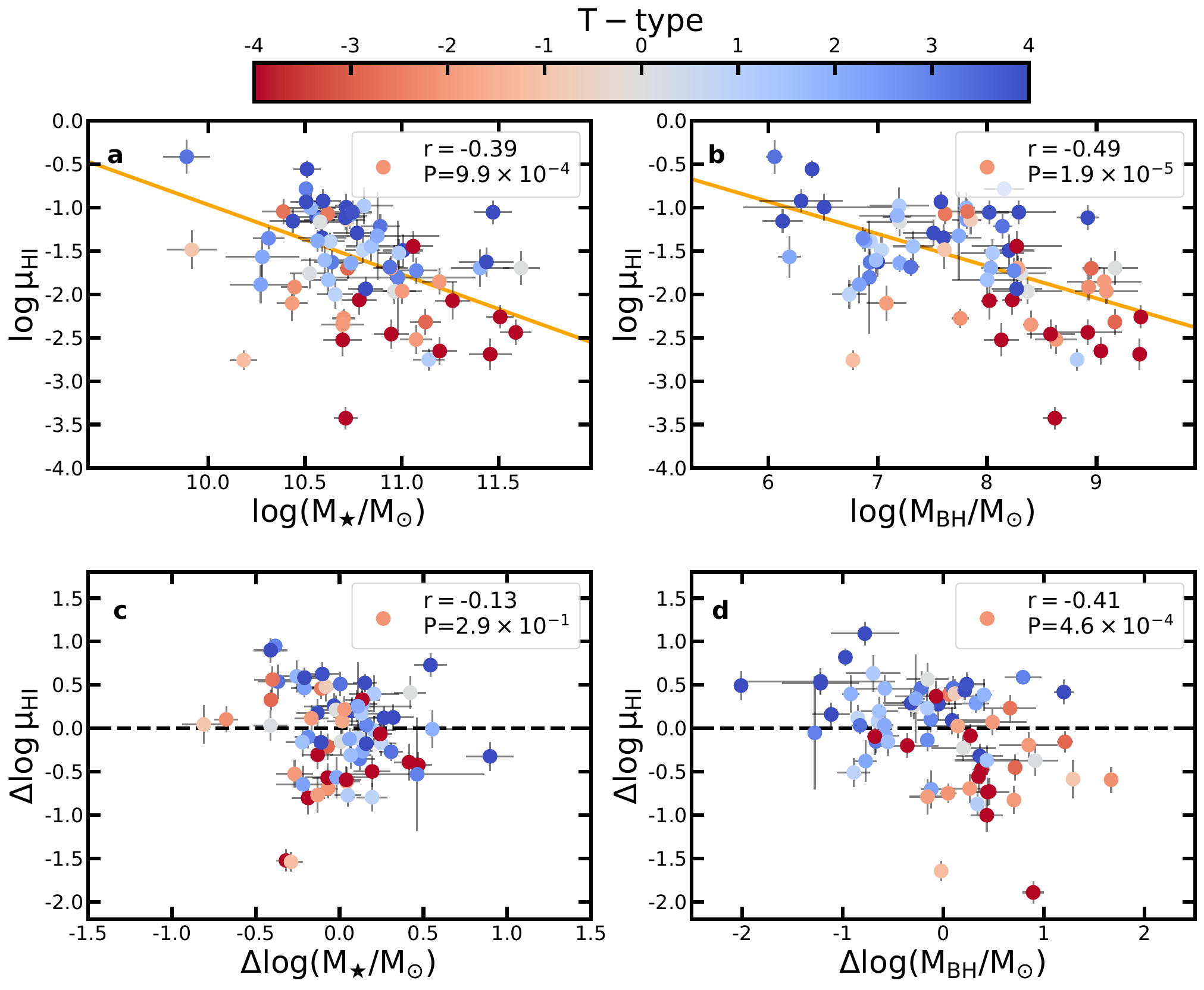}
\caption{\small{\textbf{Comparison between the relations of $\mu_{\rm HI}$ to $M_\star$ and $\mu_{\rm HI}$ to $M_{\rm BH}$ for the BH sample.} 
\textbf{a,b,} $\mu_{\rm HI}-M_{\star}$ ({\bf a}) and $\mu_{\rm HI}-M_{\rm BH}$ ({\bf b}) correlation. Galaxies are colour-coded by their morphological T types, with smaller values being more early-type and larger values more late-type morphologies. The orange lines represent the best-fitted linear relation, taking into account the uncertainties of both variables. \textbf{c,d,} Comparison of the partial correlation of $\mu_{\rm HI}-M_{\star}$ (while controlling for $M_{\rm BH}$) ({\bf c}) and $\mu_{\rm HI}-M_{\rm BH}$ (while controlling for $M_{\star}$) ({\bf d}). The x- and y-axes show the residual in $\mu_{\rm HI}$ and $M_{\star}$ after removing their dependence on $M_{\rm BH}$ in the left panel, and $M_{\rm BH}$ after removing their dependence on $M_{\star}$ in the right panel: $\Delta {\rm log} \mu_{\rm HI} = {\rm log} \mu_{\rm HI} - {\rm log} \mu_{\rm HI} (M_{\rm BH})$ and $\Delta {\rm log} M_{\star} = {\rm log} M_{\star} - {\rm log} M_{\star}(M_{\rm BH})$ in the left panel, and $\Delta {\rm log} \mu_{\rm HI} = {\rm log} \mu_{\rm HI} - {\rm log} \mu_{\rm HI} (M_{\star})$ and $\Delta {\rm log} M_{\rm BH} = {\rm log} M_{\rm BH} - {\rm log} M_{\rm BH}(M_{\star})$ in the right panel. The horizontal dashed line indicates zero correlation, that is, there is no intrinsic correlation between the two quantities. The Spearman correlation coefficients between the two corresponding variables are shown in each panel. The error bars refer to 1$\sigma$ measurement errors.}}
\label{fig1}
\end{figure*}

We define the HI gas content as the ratio of the HI mass and the stellar mass represented as $\mu_{\rm HI}$ = $M_{\rm HI}/M_{\star}$. We first examine the relationship between $\mu_{\rm HI}$ and BH masses and compare it with the $\mu_{\rm HI}-M_{\star}$ correlation in Fig.~\ref{fig1}. The correlation with $M_{\rm BH}$ is found to be more significant than the correlation with $M_{\star}$, with a Spearman correlation coefficient of $r =-0.49 (P = 10^{-4.7})$ and $r =-0.39 (P = 10^{-3.0})$, respectively. More importantly, the partial correlation between $\mu_{\rm HI}$ and $M_{\star}$ while controlling for $M_{\rm BH}$, that is, removing both their dependence on $M_{\rm BH}$, indicates that $\mu_{\rm HI}$ shows no dependence on $M_{\star}$ ($r = -0.13, P = 0.29$; Fig.~\ref{fig1} (bottom left)), whereas strong residual correlation exists between $\mu_{\rm HI}$ and $M_{\rm BH}$ while controlling for $M_{\star}$ ($r = -0.41, P = 10^{-3.3}$; Fig.~\ref{fig1} (bottom right)). 
Moreover, although the $\mu_{\rm HI}-M_{\star}$ correlation differs significantly for early- and late-type galaxies with the early-type galaxies exhibiting systematically lower $\mu_{\rm HI}$ at fixed $M_{\star}$, galaxies with different morphologies follow the same $\mu_{\rm HI}-M_{\rm BH}$ relation. This suggests that the low HI values in those early-type galaxies on the $\mu_{\rm HI}-M_{\star}$ relation are probably only a reflection that these galaxies have more massive BHs compared with late-type galaxies with similar $M_{\star}$.

Although the partial correlation between $\mu_{\rm HI}$, $M_{\rm BH}$ and $M_{\star}$ offers direct evidence that BHs play a more crucial part than $M_{\star}$ in regulating $\mu_{\rm HI}$, the heterogeneous nature of this sample makes it challenging to determine how the resulting relation could be applicable to broad galaxy populations. To validate this relation, we used a large sample of nearby galaxies with deep HI observations (Methods and Extended Data Fig.~\ref{edf1}), which comprises 474 group central galaxies with $10^{9.5} M_{\odot} < M_{\star} < 10^{11.5} M_{\odot}$ and reliable central velocity dispersion ($\sigma$) measurements. Out of this, 281 of them are detected in HI with HI upper limits available for the remaining 193 sources. $M_{\rm BH}$ for this galaxy sample is inferred from the $M_{\rm BH}-\sigma$ relation (Methods). Hereafter we will call this enlarged galaxy sample ``the galaxy sample'', and we call the sample with directly measured $M_{\rm BH}$ ``the BH sample''.

\begin{figure*}
	\centering
	\includegraphics[scale=0.4]{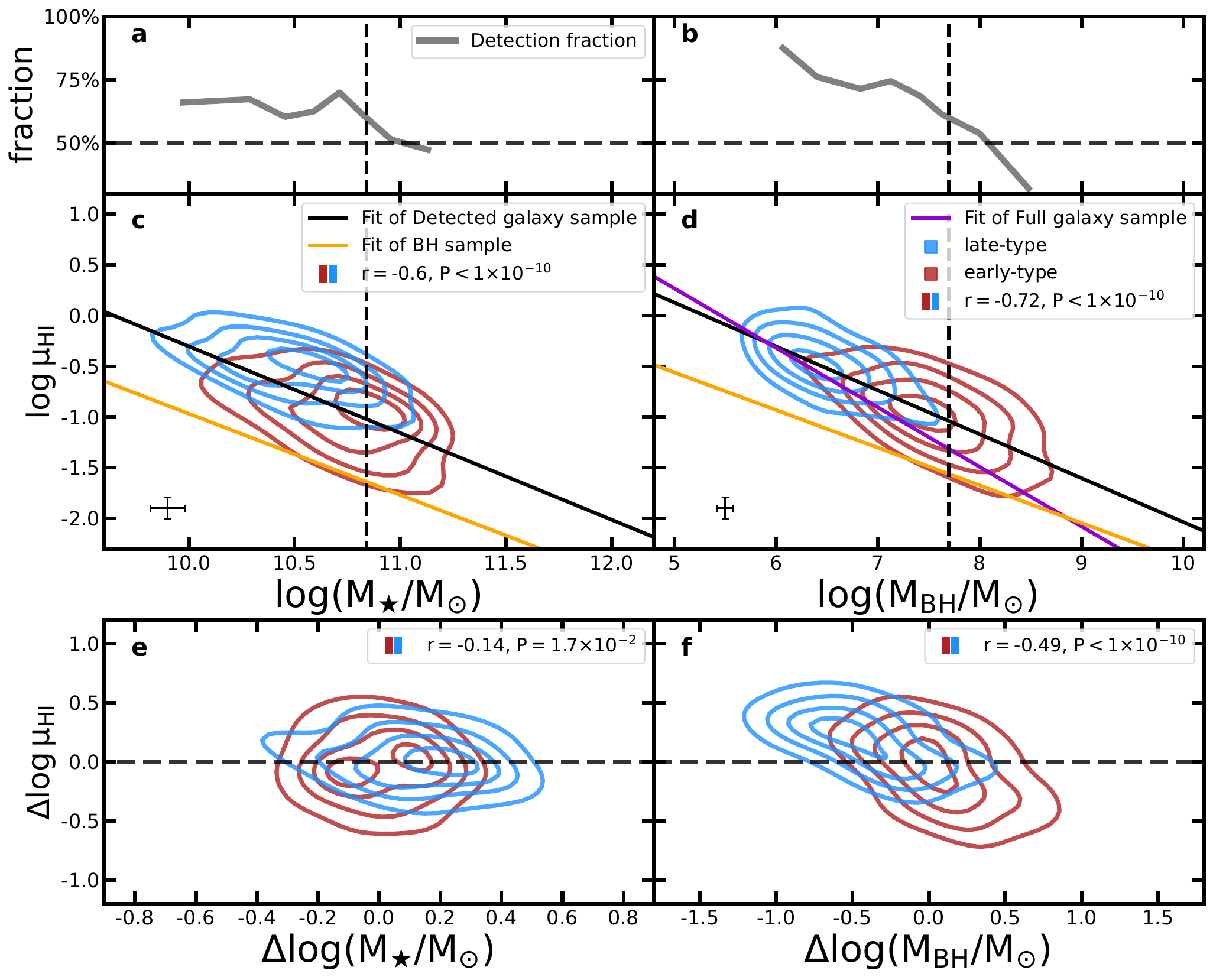}
\caption{\small{\textbf{Comparison between the relations of $\mu_{\rm HI}$ to $M_\star$ and $\mu_{\rm HI}$ to $M_{\rm BH}$ for the galaxy sample.} \textbf{a,b,} $\mu_{\rm HI}-M_{\star}$ ({\bf a}) and $\mu_{\rm HI}-M_{\rm BH}$ ({\bf b}) correlations. Galaxies are divided into early- and late-type galaxies based on their S{\'e}rsic indexes (separated at $n = 2$), which are shown in red and blue contours, respectively. The HI-detection rates of galaxies are shown as a function of stellar masses and BH masses. The vertical dashed lines indicate the position when the HI-detection fraction reaches 60\%. \textbf{c,d,} $\mu_{\rm HI}-M_{\star}$ ({\bf c}) and $\mu_{\rm HI}-M_{\rm BH}$ ({\bf d}) relations. The best-fitted relations for the HI-detected galaxy sample and the BH sample are shown by the black and orange lines, respectively. We also show the $\mu_{\rm HI}-M_{\rm BH}$ relation for the full galaxy sample with the magenta line in {\bf (d)}. \textbf{e,f,} The partial correlation between $\mu_{\rm HI}$ and $M_{\star}$ while controlling for $M_{\rm BH}$ ({\bf e}), and the partial correlation between $\mu_{\rm HI}$ and $M_{\rm BH}$ while controlling for $M_{\star}$ ({\bf f}). The corresponding Spearman coefficients are shown in each panel. The median 1$\sigma$ error bars for the galaxy sample are shown in {\bf c} and {\bf d}.}}
	\label{fig2}
\end{figure*}

The $\mu_{\rm HI}-M_{\star}$ and $\mu_{\rm HI}-M_{\rm BH}$ relations for the galaxy sample are shown in Fig.~\ref{fig2}. Both $M_{\rm BH}$ and $M_{\star}$ are found to be tightly correlated with $\mu_{\rm HI}$ with respective $r = -0.72$ and $r = -0.60$. However, the partial correlation suggests that the $\mu_{\rm HI}-M_{\star}$ correlation almost disappears (when controlling for $M_{\rm BH}$) with $r = -0.14$, compared with $r = -0.49$ for the $\mu_{\rm HI}-M_{\rm BH}$ relation (when controlling for $M_{\star}$). This further suggests that the $\mu_{\rm HI}-M_{\star}$ correlation is mostly driven by the $\mu_{\rm HI}-M_{\rm BH}$ and $M_{\star}-M_{\rm BH}$ correlations.
Similar to the BH sample, early- and late-type galaxies follow the same $\mu_{\rm HI}-M_{\rm BH}$ relation, but a different $\mu_{\rm HI}-M_{\star}$ relation.

The best-fitted $\mu_{\rm HI}-M_{\rm BH}$ relation for the HI-detected galaxy sample yields a slope of $-0.43 \pm 0.02$ (Fig.~\ref{fig2}, black line), which is steeper than that for the BH sample ($-0.37\pm0.06$; Fig.~\ref{fig2}, orange line). This is most likely driven by the selection biases in the BH sample, which are more complete and representative at large $M_{\rm BH}$ but poorly sampled at low $M_{\rm BH}$. Moreover, we also derive an inherent $\mu_{\rm HI}-M_{\rm BH}$ scaling relation (Fig.~\ref{fig2}, magenta line) encompassing both HI detections and non-detections, resulting in a steeper slope ($-0.59\pm0.19$) than that for fitting the HI detections exclusively (Extended Data Table ~\ref{etb2}).

\begin{figure*}
	\centering
	\includegraphics[scale=0.20]{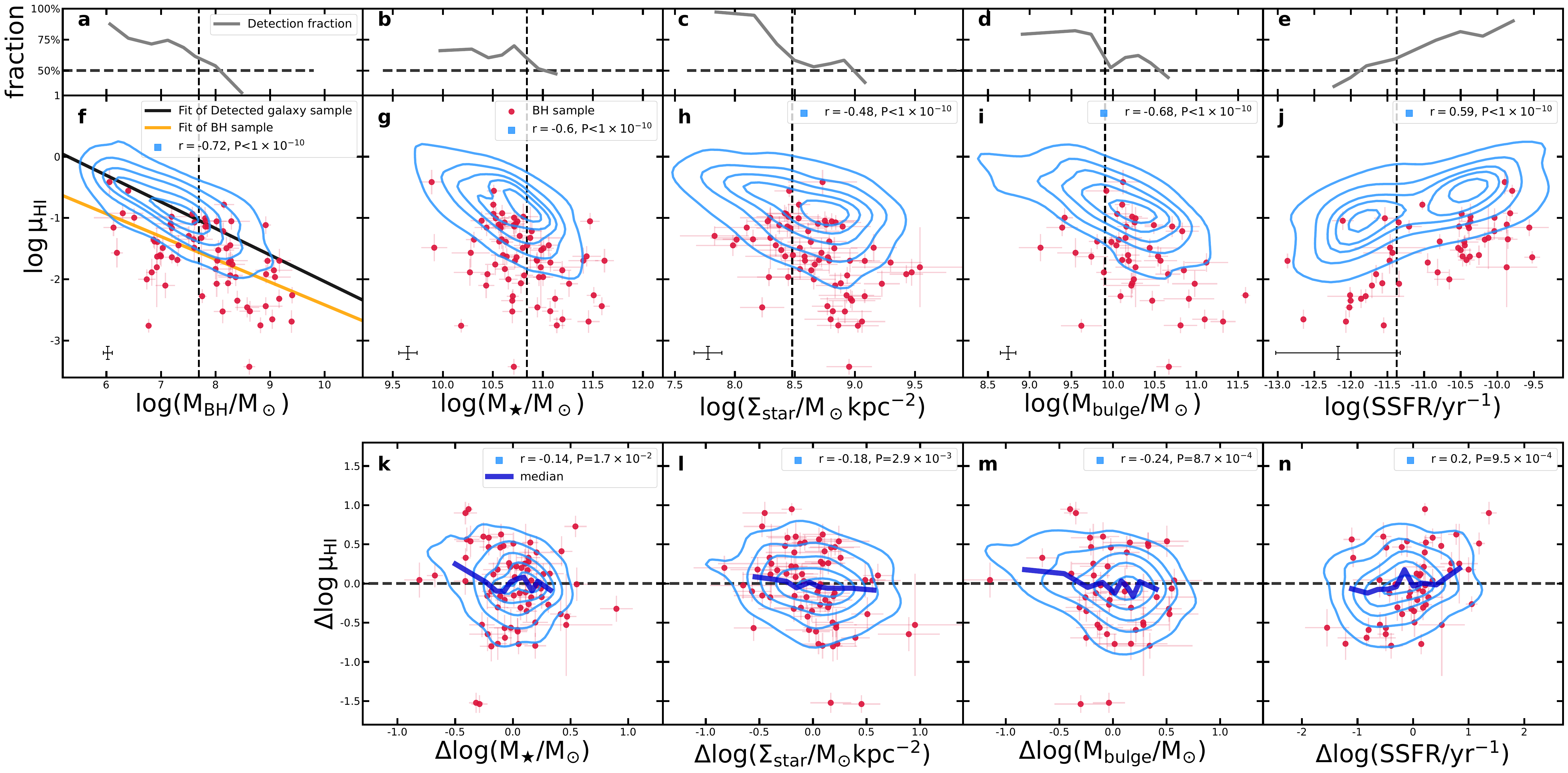}
\caption{\small{\textbf{The impact of $M_{\rm BH}$ on the correlation between $\mu_{\rm HI}$ and other main galactic parameters.} \textbf{a-e,} The HI-detection fraction along $M_{\rm BH}$ ({\bf a}) and some other main physical parameters of galaxies, including $M_{\star}$ (\textbf{b}), $\Sigma_{\rm star}$ (\textbf{c}), $M_{\rm bulge}$ (\textbf{d}) and $\rm{SSFR}$ (\textbf{e}). The vertical dashed lines indicate the position at which the HI-detection rates hit 60\%. \textbf{f-j,}  The relation between the parameters $M_{\rm BH}$ (\textbf{f}), $M_{\star}$ (\textbf{g}), $\Sigma_{\rm star}$ (\textbf{h}), $M_{\rm bulge}$ (\textbf{i}) and $\rm{SSFR}$ (\textbf{j}) and $\mu_{\rm HI}$. The contours denote the distribution of the HI-detected galaxy sample, whereas the filled red circles denote the BH sample with 1$\sigma$ error bars. The best-fitted $\mu_{\rm HI}-M_{\rm BH}$ relations for the HI-detected galaxy sample and the BH sample are shown in \textbf{f} by the black and orange lines, respectively. The median 1$\sigma$ error bars for the galaxy sample are shown. \textbf{k-n,} The relation between the residual in $\mu_{\rm HI}$ and the residual in other galactic parameters after removing their dependence on $M_{\rm BH}$: $\Delta {\rm log}\mu_{\rm HI} = {\rm log}\mu_{\rm HI}- {\rm log}\mu_{\rm HI}(M_{\rm BH})$ and $\Delta {\rm \log} X = {\rm \log}X- {\rm \log} X(M_{\rm BH})$ with $X$ representing $M_{\star}$ (\textbf{k}), $\Sigma_{\rm star}$ (\textbf{l}), $M_{\rm bulge}$ (\textbf{m}) and ${\rm SSFR}$ (\textbf{n}), and $\mu_{\rm HI}(M_{\rm BH})$ and $X(M_{\rm BH})$ derived from their best-fitted relation with $M_{\rm BH}$ (Extended Data Fig.~\ref{edf2}). The solid medium-blue lines in \textbf{k-n} show the running median of the residuals in $\mu_{\rm HI}$. The Spearman correlation coefficients for the HI-detected galaxy sample between the corresponding $x$ and $y$ variables are shown in \textbf{f-n}. }}
\label{fig3}
\end{figure*}

Next we explore further the correlations between $\mu_{\rm HI}$ and other main galactic parameters\cite{Saintonge:2022}, including stellar surface densities ($\Sigma_{\rm star}$), bulge masses ($M_{\rm bulge}$), and specific star formation rates (SSFR), to determine whether $M_{\rm BH}$ is the key parameter in determining $\mu_{\rm HI}$ in galaxies. Figure ~\ref{fig3} compares the correlation among $\mu_{\rm HI}$, $M_{\star}$, $M_{\rm BH}$, $\Sigma_{\rm star}$, $M_{\rm bulge}$, and SSFR for the HI-detected galaxy sample. Although significant correlations exist between $\mu_{\rm HI}$ and all these parameters, after removing their dependence on $M_{\rm BH}$, all the correlations almost disappear, with negligible Spearman coefficients and zero running medians. We also verify this by using the inherent $\mu_{\rm HI}-M_{\rm BH}$ relation derived for the full sample in Extended Data Fig.~\ref{edf3}, yielding consistent results.

Given that $M_{\rm BH}$, $M_{\star}$, $\Sigma_{\rm star}$ and $M_{\rm bulge}$ are all highly correlated, as a further test on the fundamental role of $M_{\rm BH}$ in driving the correlation with $\mu_{\rm HI}$, we conduct a partial least squares regression between $\mu_{\rm HI}$ and the parameter set of $M_{\rm BH}$, $M_{\star}$, $\Sigma_{\rm star}$ and $M_{\rm bulge}$ for the HI-detected galaxy sample and the BH sample, which shows that $M_{\rm BH}$ is the most significant predictor parameter of $\mu_{\rm HI}$ (Methods).

As $M_{\rm BH}$ is proportional to the integrated energy of BHs across their accretion history\cite{Silk:1998,Bluck:2022}, our findings offer observational evidence that the accumulated energy from BHs is vital in regulating the accretion and/or cooling of cool gas in galaxies. 
The immense energy released from the accretion of SMBHs in massive galaxies is known to be at least comparable to the binding energy of host galaxies\cite{Silk:1998,Bluck:2011,Maiolino:2012}. This energy is thought to significantly affect the accretion of gas onto the galaxy and the cooling of the circumgalactic medium (CGM) and ISM. As $M_{\star}$ is closely linked with the inner halo binding energy (total binding energies within effective radii of galaxies)\cite{ShiY:2021b}, $M_{\star} \propto E_{\rm b}^{\beta}$, the $\mu_{\rm HI}-M_{\rm BH}$ relation means $M_{\rm HI} \propto M_{\star} M_{\rm BH}^{-\alpha} \propto E_{\rm b}^{\beta}M_{\rm BH}^{-\alpha}$, where $E_{\rm b}$ represents the binding energy of the inner dark matter halo. At the stellar mass range probed by our BH sample, $\beta \sim 0.6$, which is close to the value of $\alpha$, yielding $M_{\rm HI} \propto (E_{\rm b}/M_{\rm BH})^{\alpha}$ with $\alpha \sim 0.6$.

\begin{figure*}
	\centering
	\includegraphics[scale=0.08]{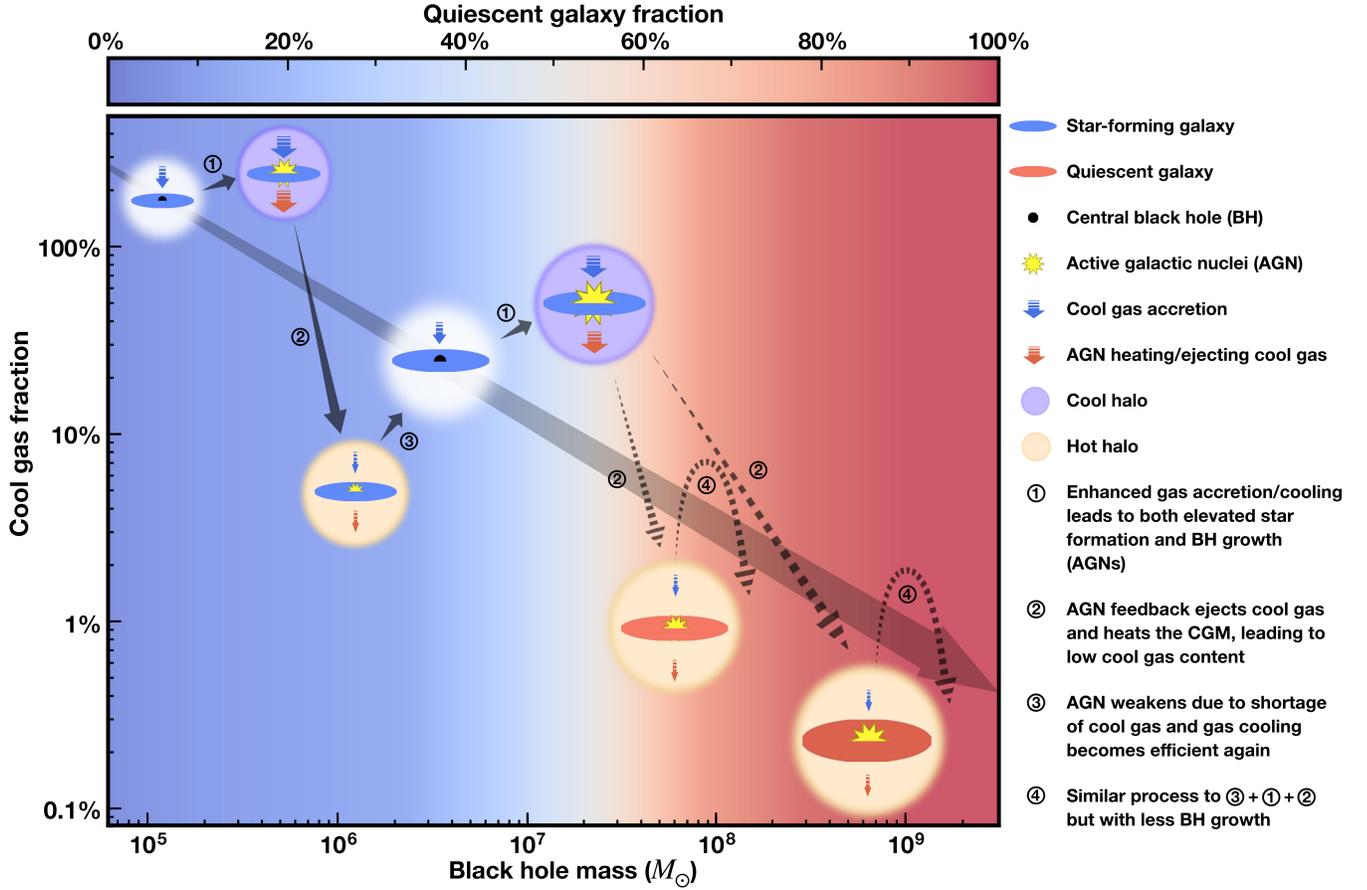}
	\caption{\small{\textbf{Schematic of the proposed scenario on how BHs regulate cool gas content in galaxies.} The large arrow indicates the $\mu_{\rm HI}-M_{\rm BH}$ correlation. The background colour scale indicates the quiescent galaxy fraction as a function of $M_{\rm BH}$, which shows a sharp increase at $M_{\rm BH} \gtrsim 10^{7.5} M_{\odot}$ (Methods and Extended Data Fig.~\ref{edf5}), corresponding to $\mu_{\rm HI} < 10\%$. At fixed $M_{\rm BH}$, galaxies could maintain their $\mu_{\rm HI}$ at a certain level determined by the relative strength of the inner halo binding energy and $M_{\rm BH}$. Once gas accretion is enhanced onto galaxies (and their BHs), which increases $\mu_{\rm HI}$, $M_{\rm BH}$ will also grow and release additional heating energy that prevents further gas cooling or accretion. This will bring down $\mu_{\rm HI}$ together with increasing $M_{\star}$ by star formation and reach a new balance at higher $M_{\rm BH}$. Although the same process takes place in both SFGs and quiescent galaxies, the growth of $M_{\rm BH}$ or $M_{\star}$ should be much less significant in quiescent galaxies than in SFGs, and the large range of $M_{\rm BH}$ among quiescent galaxies ($M_{\rm BH} \sim 10^{7-10} M_{\odot}$) is probably inherited from their different star-forming progenitors when they were quenched.}}
\label{fig4}
\end{figure*}

The analysis above indicates that the HI mass in galaxies is determined by the relative strength between the binding energy of the halo and the energy released from BHs ($E_{\rm BH} \propto M_{\rm BH}$). The binding energy of the halo determines how much gas can be accreted onto the dark matter halo, whereas the energy from BHs ejects or heats up the gas, preventing it from further cooling. The contest between the two determines how much accreted gas can be eventually cooled and settled down onto the central galaxies. For such a mechanism to be effective, a negative feedback loop involving gas accretion or cooling and BH accretion or feedback would be required\cite{Fabian:2012,Bluck:2022}. The fact that the accreted cool gas could feed both star formation and BH accretion makes this possible. When gas accretion or cooling is elevated, stronger BH accretion is also triggered, resulting in more energy ejected into the ISM and CGM, which inhibits further cooling or accretion of the cool gas. This eventually brings down the cool gas content (and also the BH accretion rates). Conversely, a lower cool gas content would generally lead to weaker BH accretion with less energy ejection into the ISM and CGM, which will facilitate further cool gas accretion or cooling and increase $\mu_{\rm HI}$ until it reaches the average relation. The same physical process applies to both star-forming galaxies (SFGs) and quiescent galaxies. The difference is that although both $M_{\rm BH}$ and $M_{\star}$ of SFGs can grow substantially through this process, most quiescent galaxies will probably maintain their $M_{\rm BH}$ and $M_{\star}$ when they are quenched because of their overall low BH accretion rates and star formation rates. This scenario is shown in Fig.~\ref{fig4}.

Under this scenario, the correlation between the total gas fraction ($\mu_{\rm HI + H_2}$) and $M_{\rm BH}$ is expected to be even tighter than the $\mu_{\rm HI}-M_{\rm BH}$ relation. This is because gas cooling from the CGM will probably first cool as HI gas and only later become molecular gas that hosts star formation. In other words, HI gas probes only one phase of the cold gas, whereas AGN feedback should affect the cooling of both atomic and molecular gas in galaxies. This is probably the case. Although the sample with both HI and CO measurements is small, a reduced scatter for the $\mu_{\rm HI + H_2}-M_{\rm BH}$ relation is found compared with the $\mu_{\rm HI}-M_{\rm BH}$ relation (Methods and Extended Data Fig.~\ref{edf4}). Apart from the small sample size, most of the galaxies with both HI and CO measurements are SFGs. Future studies with much larger and more representative galaxy samples with both HI and CO measurements will be needed to fully verify the $\mu_{\rm HI + H_2}-M_{\rm BH}$ relation.

As cool gas is the material of star formation, these findings also shed critical light on the intimate connection between the presence of massive BHs and the quiescence of galaxies. It explains well why most quiescent galaxies are present only at $M_{\rm BH} \gtrsim 10^{7.5}M_{\odot}$ \cite{Terrazas:2016b,Terrazas:2017,Piotrowska:2022,Bluck:2022} (Extended Data Fig.~\ref{edf5}), corresponding to a low level of cool gas content ($\lesssim10\%$), hence minimal star formation rates. The proposed mechanism reconciles the discrepancy between the absence of strong instantaneous negative AGN feedback and the tight correlation between $M_{\rm BH}$ with galaxy quiescence. It is also consistent with empirical models indicating that the contest between dark matter halos and BHs governs the quenching of star formation in galaxies based on various observed galactic scaling relations\cite{ChenZ:2020}.

Although current studies have been confined to galaxies in the local Universe, the strong correlation across all redshifts between the quiescence of a galaxy and a prominent bulge, a high central stellar density or high central gravitational potential\cite{Bell:2012,WangT:2012,FangJ:2013,Bluck:2023,Bluck:2024}, all of which suggest a large BH, implies that the same scenario may be applied to galaxies at high redshifts as well. Next-generation facilities, such as the Square Kilometer Array and the Next Generation Very Large Array, would be required to confirm this.

\clearpage

\vspace{3pt}
\noindent\rule{\linewidth}{0.4pt}

\end{bibunit}

\clearpage
\appendix

\begin{bibunit}

\noindent\textbf{\large Methods}
\section{Cosmology}
We adopted a Chabrier initial mass function (IMF)\cite{Chabrier:2003} to estimate star formation rate (SFR) and assumed cosmological parameters of $H_{0}$ = 70 km s$^{-1}$ Mpc$^{-1}$, $\Omega_{M}$ = 0.3, and $\Omega_{\Lambda}$ = 0.7. 

\section{Sample selection:} 

\subsection{The BH sample}
The sample for galaxies with directly measured BH masses is primarily from ref.\cite{Terrazas:2017}, which includes 91 central galaxies collected from refs.\cite{Saglia:2016,vandenBosch:2016,Kormendy:2013}. We excluded 18 sources with BH masses measured with reverberation mapping  and kept only those masses measured with dynamical methods. We then added another 63 galaxies with measured BH masses from recent literature, which were matched with the group catalogue\cite{LuY:2016} of nearby galaxies to select only central galaxies. We obtained the HI flux densities and masses of this sample by crossmatching with the nearby galaxy database, HyperLeda\cite{Makarov:2014}. Our final sample includes 69 central galaxies with 41 from ref.\cite{Terrazas:2017} and the remaining from the compilation of recent literature. In Extended data table \ref{etb1}, we list the basic properties of our BH sample.

\subsection{The galaxy sample}
The sample for galaxies with HI measurements and indirect BH mass measurements are from the extended GALEX Arecibo SDSS Survey (xGASS; ref.\cite{Catinella:2018}) and HI-MaNGA programme\cite{Masters:2019,Stark:2021}, which include HI observations towards a representative sample of about 1,200 and 6,000 galaxies with $10^{9} M_{\odot} < M_{\star} < 10^{11.5}M_{\odot}$, respectively. 
The depth of the survey also allows for  stringent constraints on the upper limits for the HI non-detections , enabling a comprehensive assessment of $\mu_{\rm HI}$ for the entire sample. 
We limited the redshift ${\rm z}<0.035$ to ensure high HI-detection rates even at the highest stellar masses and BH masses. We selected only group central galaxies, which include at least one satellite galaxy in their groups, based on the crossmatch with the group catalogue{{\cite{Abdurro:2022,Yang:2007,Janowiecki:2017}}}. Isolated central galaxies lacking any satellites in their groups are discarded because they may have probably suffered from additional environmental effects\cite{WangK:2023}. 
We derived the BH masses for the xGASS and HI-MaNGA sample with their velocity dispersion~\cite{vandenBosch:2016} from SDSS DR17\cite{Abdurro:2022} ($\sigma_{\rm SDSS}$, and we require $\sigma_{\rm SDSS}\geq 70~{\rm km/s}$): 
\begin{equation}
    \log \left(\frac{M_{\rm BH}}{M_{\odot}}\right)= (8.32\pm0.04) + (5.35\pm0.23) \log\left(\frac{\sigma_{\rm SDSS}}{200~\rm km~s^{-1}}\right).
    \label{eq:MS}
\end{equation}

\section{Physical parameters of the BH and galaxy sample:}

\subsection{Stellar masses}
The stellar masses for the galaxy sample are taken from MPA-JHU catalogue\cite{Kauffmann:2003,Brinchmann:2004}, which are derived from SED fitting based on SDSS data. For the BH sample, because most of them lack the same photometric coverage as the galaxy sample, we derive their stellar masses from their K-band luminosity and velocity dispersion-dependent K-band mass-to-light ratio following ref.\cite{vandenBosch:2016}:
\begin{equation}
    M_{\star}/L_{\rm K} = 0.1 \sigma_{\rm e}^{0.45}.
    \label{eq2}
\end{equation}
As an accurate determination of $\sigma_{e}$ is {{not available for all}} galaxies, we derived  $\sigma_{\rm e}$ for the full BH sample from the tight correlation in ref.\cite{vandenBosch:2016}:
\begin{equation}
\begin{split}
        \log\left(\frac{\sigma_{\rm e}}{\rm km s^{-1}}\right)=(2.11\pm0.01)+ & (0.71\pm0.03)\log\left(\frac{L_{\rm K}}{10^{11}L_{\odot}}\right)\\
    + & (-0.72\pm0.05)\log\left(\frac{R_{\rm e}}{5~{\rm kpc}}\right).
\label{eq3}
\end{split}
\end{equation}
To explore whether there are systematic differences between the two methods, we compare the stellar masses of the galaxy sample taken from the MPA-JHU catalogue and those deived from equation~(\ref{eq2}). A median mass difference 0.32\,dex is found between the two methods (Extended Data Fig.~\ref{edf6}), which may be attributed to the tilt from the fundamental plane beyond the mass-to-light ratio, for example, the dark matter component in the effective radius. We corrected these systematic mass differences for the BH sample to match that of the galaxy sample.

\subsection{HI fraction and upper limits}
The HI-detection limit depends not only on the sensitivity but also on the width of the HI line. To obtain more realistic upper limits, we first derived the expected HI line width for each HI non-detection. The width of the HI line indicates the circular velocity of the host galaxy, which should be proportional to the stellar masses. We explored this using the HI detections from the xGASS and HI-MaNGA sample. Extended Data Figure~\ref{edf1} shows the relation between $M_{\star}$ and the observed line width, as well as $M_{\star}$ and inclination-corrected line width. It indicates that the inclination-corrected line width is tightly correlated with $M_{\star}$, which is further used to derive expected line width for the HI non-detections. Combining the sensitivity of the HI observations and the expected line width, we derived the upper limits for all the HI non-detections in our BH and galaxy samples.

\subsection{Morphology}
For BH sample, the morphology indicator $T$ is obtained from the HyperLEDA database\cite{Makarov:2014}. It can be a non-integer because for most objects the final $T$ is averaged over various estimates available in the literature. For the galaxy sample, we classified them in to the early types and late types based on the S$\rm\acute{e}$rsic index (from NASA-Sloan Atlas catalog; NSA: Blanton M. \href{http://www.nsatlas.org}{{http://www.nsatlas.org}}) larger or smaller than 2.

\subsection{Star formation rates} The specific star formation rates (SSFR) of galaxy sample are from the MPA-JHU catalogue based on ref.\cite{Brinchmann:2004}. The SSFR for the BH sample is taken from the original reference.

\subsection{Bulge masses}
The bulge information is from refs.\cite{Bohn:2020,Simard:2011} for the BH sample and galaxy sample, respectively. {{More specifically, we calculate the bulge mass for galaxy sample using r-band $B/T$. }}

\subsection{Stellar mass surface density}
{{We calculated the K-band effective radius for both the BH and galaxy sample according to ref\cite{vandenBosch:2016}}: $\log R_{\rm e}=1.16\log R_{\rm K\_R\_EFF}+0.23\log q_{\rm K\_BA}$, where $R_{\rm e}$ is the corrected apparent effective size, $R_{\rm K\_R\_EFF}$ and $q_{\rm K\_BA}$ are K-band apparent effective radius and K-band axis ratio from 2MASS. After converting the apparent sizes to the physical sizes, the stellar mass surface density was derived as: $\Sigma_{\rm star}=M_\star/(2\pi R_{\rm e}^2)$.}

\subsection{$\rm \bf{H_2}$ masses}
{We collected $\rm H_2$ masses from xCOLD GASS survey\cite{Saintonge:2017} and ref.\cite{Wylezalek:2022} for xGASS and MaNGA galaxies, respectively. We acknowledge that at least in the nearby Universe, the molecular-to-atomic gas mass ratio increases only weakly with stellar masses and remains relatively low over a wide stellar mass range, with $R\equiv M_{\rm H_2}/M_{\rm HI}\sim 10-20\%$ at $10^9 M_{\odot} < M_{\star} < 10^{11.5}M_{\odot}$. We calculate the total gas fractions as $\mu_{\rm HI+H_{2}} = ( M_{\rm HI} + M_{\rm H_{2}})/M_{\star}$. For central galaxies (isolated centrals plus group centrals), we compare the $M_{\rm BH}-\mu_{\rm HI}$ and $M_{\rm BH}-\mu_{\rm HI+H_{2}}$ relation in Extended Data Fig.~\ref{edf4}. The $M_{\rm BH}-\mu_{\rm HI+H_{2}}$ relation exhibits a stronger correlation with smaller scatter than the $M_{\rm BH}-\mu_{\rm HI}$ relation. We acknowledge that, based on molecular hydrogen gas content traced through dust extinction, previous studies reveal a $M_{\rm BH}-\mu_{\rm H_2}$ correlation\cite{Piotrowska:2022}. Future studies with more direct measurements of molecular hydrogen gas for large samples will be needed to examine in detail whether $M_{\rm BH}$ also plays a fundamental role in regulating the molecular gas content in galaxies.}

\section{Quiescent fraction}
{To estimate the quiescent fraction at different $M_{\rm BH}$, we selected galaxies from the MPA-JHU catalog of SDSS galaxies with the same criteria as the galaxy sample, except that we limited velocity dispersion to greater than $30\rm \,km/s$ to cover broader $M_{\rm BH}$ and we made no constraints on the HI detection. We classified the sample galaxies into star-forming and quiescent ones, separated at SSFR$=-11$. In each $M_{\rm BH}$ bin, the quiescent fraction was calculated as the ratio between the number of quiescent galaxies and that of all galaxies. The result is shown in Extended Data Fig.~\ref{edf5}, which is consistent with that of previous work\cite{Terrazas:2020,Bluck:2023}.}

\section{Linear Least-squares approximation}
We implemented linear regression for the BH sample and the galaxy sample using Python package {LTS\_LINEFIT} introduced in ref.\cite{Cappellari:2013}, which is insensitive to outliers and can give the intrinsic scatter around the linear relation with corresponding errors of the fitted parameters. 

\section{Linear fitting including upper limits}
To incorporate both detections and upper limits in the galaxy sample, we applied the Kaplan-Meier (KM) non-parametric estimator to derive the cumulative distribution function at different $M_{\rm BH}$ bins (with Python package {Reliability}~\cite{Reid:2022}), and performed 10,000 random draws from the cumulative distribution function at each bin to fit the relation between $\mu_{\rm HI}$ and $M_{\rm BH}$. The linear relation and its corresponding errors are taken as the best fitting and standard deviations of these fittings (Extended Data Table~\ref{etb2}).
The non-detection rate of HI is relatively low across most of the $M_{\rm BH}$ range and becomes significant only for galaxies with the most massive BHs (reaching about $50\%$ at $M_{\rm BH} > 10^{8}~M_{\odot}$). 

\section{Partial least square regression} To derive the most significant physical parameters in determining $\mu_{\rm HI}$ statistically, we used Python package Scikit-learn\cite{pedregosa:2011} with partial least square (PLS) Regression function, which uses a non-linear iterative partial least squares (NIPALS)\cite{Lindgren1993} algorithm. {The PLS algorithm generalizes a few latent variables (or principal components) that summarize the variance of independent variables, which
is used to find the fundamental relation between a set of independent and dependent variables.} It has advantages in regression among highly-correlated predictor variables. It calculates the linear combinations of the original predictor datasets (latent variables) and the response datasets with maximal covariance, then fits the regression between the projected datasets and returns the model:
\begin{equation}
    Y=XB+F,
\end{equation}
 where $X$ and $Y$ are predictor and response datasets, $B$ is the matrix of regression coefficients and $F$ is the intercept matrix.
 
We constructed the $X$ and $Y$ matrix as the set of $M_{\rm BH}$, $M_{\star}$, $\Sigma_{\rm star}$, $M_{\rm bulge}$ and the set of $\mu_{\rm HI}$. For the BH and galaxy samples, this returns the sample size of 45 and {{189}}, respectively. The optimal number of latent variables (linear combinations of predictor variables) in PLS Regression is determined by the minimum of mean squared error from cross-validation (using function cross\_val\_predict in Scikit-learn) at each number of components. We find that the optimal number of latent variables for both the BH and the galaxy sample converges to one.
Further increasing the number of latent variables only yields a few percentage changes in the mean squared errors, and $M_{\rm BH}$ remains the most significant predictor parameter. Following appendix B in refs.\cite{Oh2022}, the variance contribution from different parameters to $\mu_{\rm HI}$ is decomposed as:
\begin{equation}
    {\rm Var}(Y)=\sum_{i=1}^{4}{\rm Var}(X_i B_i)+{\rm Var}(F),
\end{equation}
where Var is a measure of the spread of a distribution. The portion of each parameter variance is shown in the last column of the Extended Data Table~\ref{etb3}, which clearly shows that $M_{\rm BH}$ dominates the variance. Further increasing the number of latent variables results only in a few percentage changes in the mean squared errors, and $M_{\rm BH}$ remains the most significant predictor parameter.

\vspace{3pt}
\noindent\rule{\linewidth}{0.4pt}
\vspace{3pt}
\let\oldthebibliography=\thebibliography
\let\oldendthebibliography=\endthebibliography
\renewenvironment{thebibliography}[1]{%
    \oldthebibliography{#1}%
    \setcounter{enumiv}{30} 
}{\oldendthebibliography}

\begin{addendum}
 \item[Acknowledgements] T.W. acknowledges support by the National Natural Science Foundation of China (NSFC, Project No. 12173017 and Key Project No. 12141301), and the China Manned Space Project (No. CMS-CSST-2021-A07). LCH was supported by NSFC (11991052, 12233001), the National Key R\&D Program of China (2022YFF0503401), and the China Manned Space Project (CMS-CSST-2021-A04, CMS-CSST-2021-A06). Z.Y.Z acknowledges the support of NSFC under grants 12173016, 12041305, the Program for Innovative Talents, Entrepreneur in Jiangsu, and the science research grants from the China Manned Space Project, CMS-CSST-2021-A08.
 Q.G. was supported by the NSFC ( 12192222, 12192220, and 12121003).
 F.Y. was supported in part by NSFC (12133008, 12192220, and 12192223).
 \item[Author Information] The authors declare no competing interests. 

 \item[Author Contributions] 
 T.W. initiated the study, led the first discoveries and authored the majority of the text. K.X. enlarged the sample and consolidated the results with more in-depth analysis under the supervision of T.W.. Y-X.W. helps construct the initial sample. Y.S., D.E., L.H., Z.Z., Q.G., Y-J.W., C.S., F.Y., X.X. and K.W. contributed to the overall interpretation of the results and various aspects of the analysis.

 \item[Data availability] 
 All data used in this paper are publicly available and we summarize the key physical parameters of the BH sample in Extended Data Table.

 \item[Code availability] 
 All codes used in the paper are publicly available.
\end{addendum}

\renewcommand{\arraystretch}{0.6}
\renewcommand{\baselinestretch}{1}

\newpage
\onecolumn
\noindent\textbf{\large Extended Data}
\renewcommand\thefigure{\arabic{figure}}
\setcounter{figure}{0}
\renewcommand{\tablename}{Extended Data Table}
\begin{small}
\begin{ThreePartTable}
\begin{longtable}{cccccccc}
\caption{\textbf{BH sample properties}}
\label{etb1}\\
\vspace{-0.5cm}
\tabcolsep=0.3cm \\
\hline \hline 
Name & log $R_{\rm e}$ & Distance & log $M_{\rm BH}$ & log $M_{\rm star}$ & log $M_{\rm HI}$\tnote{*} & log $M_{\rm bulge}$\tnote{**}  \\ 
~ & (log kpc) & (Mpc)& (log $M_{\odot}$) & (log $M_{\odot}$) & (log $M_{\odot}$) & (log $M_{\odot}$)\\ 
\midrule
\vspace{-0.6cm}
\endfirsthead\\
\multicolumn{7}{c}{(continued)}\\
\hline \hline 
 Name & log $R_{\rm e}$ & Distance & log $M_{\rm BH}$ & log $M_{\rm star}$ & log $M_{\rm HI}$\tnote{*} & log $M_{\rm bulge}$\tnote{**}  \\ 
~ & (log kpc) & (Mpc)& (log $M_{\odot}$) & (log $M_{\odot}$) & (log $M_{\odot}$) & (log $M_{\odot}$)\\ 
\midrule
\vspace{-0.3cm}
\endhead\\
Centaurus-A & $0.41 \pm 0.06$~\cite{vandenBosch:2016}  & $ 3.62 	       $~\cite{Kormendy:2013} 		& $7.76 ^{+0.08 }_{-0.08 }$~\cite{Kormendy:2013}		&  $ 11.02 \pm	0.06  $  &  $8.42 \pm 0.09 $  &   $ 10.09 \pm 0.15         $ \\
Circinus    & $0.18 \pm 0.13$~\cite{vandenBosch:2016}  & $ 2.82  \pm  0.47 $~\cite{Terrazas:2016b} & $6.06 ^{+0.08 }_{-0.08 }$~\cite{Kormendy:2013}		&  $ 10.21 \pm	0.12  $  &  $9.47 \pm 0.16 $  &   $ 10.12 \pm 0.20         $ \\
IC1459      & $0.80 \pm 0.08$~\cite{vandenBosch:2016}  & $ 28.92 \pm  3.74 $~\cite{Terrazas:2016b} & $9.39 ^{+0.08 }_{-0.03 }$~\cite{Kormendy:2013}		&  $ 11.78 \pm	0.11  $  &  $8.77 \pm 0.15 $  &   $ 11.32 \pm 0.15         $ \\
IC1481      & $0.50 \pm 0.06$~\cite{vandenBosch:2016}  & $ 89.90           $~\cite{Kormendy:2013} 		& $7.17 ^{+0.13 }_{-0.13 }$~\cite{Kormendy:2013}		&  $ 10.88 \pm	0.10  $  &  $9.45 \pm 0.10 $  &   $ 		  ...               $ \\
M31         & $0.59 \pm 0.04$						   & $ 0.77  \pm  0.03 $~\cite{Terrazas:2016b} & $8.16 ^{+0.28 }_{-0.09 }$~\cite{Kormendy:2013}		&  $ 10.83 \pm	0.04  $  &  $9.72 \pm 0.07 $  &   $ 10.11 \pm 0.09         $ \\
M66         & $0.63 \pm 0.07$~\cite{vandenBosch:2016}  & $ 10.10 \pm  1.10 $~\cite{vandenBosch:2016} 	& $6.93 ^{+0.05 }_{-0.05 }$~\cite{vandenBosch:2016}   	&  $ 10.96 \pm	0.09  $  &  $9.01 \pm 0.10 $  &   $ 9.74  \pm	0.20        $ \\ 
M81         & $0.44 \pm 0.11$~\cite{vandenBosch:2016}  & $ 3.60  \pm  0.13 $~\cite{Terrazas:2016b} & $7.81 ^{+0.17 }_{-0.10 }$~\cite{Kormendy:2013}   		&  $ 10.90 \pm	0.08  $  &  $9.44 \pm 0.07 $  &   $ 10.16 \pm 0.11         $ \\
NGC0315     & $1.01 \pm 0.05$~\cite{vandenBosch:2016}  & $ 57.70 \pm  5.80 $~\cite{vandenBosch:2016} 	& $8.92 ^{+0.31 }_{-0.31 }$~\cite{vandenBosch:2016}   	&  $ 11.91 \pm	0.08  $  &  $9.15 \pm 0.12 $  &   $ 		  ...               $ \\
NGC0613     & $0.82 \pm 0.11$~\cite{vandenBosch:2016}  & $ 15.40 \pm  1.50 $~\cite{vandenBosch:2016} 	& $7.60 ^{+0.35 }_{-0.35 }$~\cite{vandenBosch:2016}   	&  $ 10.91 \pm	0.12  $  &  $9.24 \pm 0.10 $  &   $ 		  ...               $ \\
NGC1023     & $0.49 \pm 0.10$~\cite{vandenBosch:2016}  & $ 10.81 \pm  0.80 $~\cite{Terrazas:2016b} & $7.62 ^{+0.05 }_{-0.04 }$~\cite{Kormendy:2013}   		&  $ 10.94 \pm	0.07  $  &  $9.54 \pm 0.10 $  &   $ 10.26 \pm 0.15         $ \\
NGC1068     & $0.32 \pm 0.20$~\cite{vandenBosch:2016}  & $ 15.90 \pm  9.41 $~\cite{Terrazas:2016b} & $6.92 ^{+0.02 }_{-0.02 }$~\cite{Kormendy:2013}   		&  $ 11.30 \pm	0.40  $  &  $9.18 \pm 0.52 $  &   $ 10.27 \pm 0.24         $ \\
NGC1097     & $0.74 \pm 0.12$~\cite{vandenBosch:2016}  & $ 14.50 \pm  1.50 $~\cite{vandenBosch:2016} 	& $8.14 ^{+0.09 }_{-0.09 }$~\cite{vandenBosch:2016}   	&  $ 11.21 \pm	0.10  $  &  $9.67 \pm 0.10 $  &   $ 10.83 \pm 0.20         $ \\
NGC1194     & $0.74 \pm 0.06$~\cite{vandenBosch:2016}  & $ 57.98 \pm  6.30 $~\cite{Terrazas:2016b} & $7.85 ^{+0.02 }_{-0.02 }$~\cite{Kormendy:2013}   		&  $ 11.04 \pm	0.10  $  &  $9.58 \pm 0.14 $  &   $ 10.71 \pm 0.33         $ \\
NGC1320     & $0.60 \pm 0.08$~\cite{vandenBosch:2016}  & $ 49.10 \pm  4.90 $~\cite{vandenBosch:2016} 	& $6.74 ^{+0.16 }_{-0.16 }$~\cite{vandenBosch:2016}   	&  $ 10.98 \pm	0.09  $  &  $8.65 \pm 0.14 $  &   $ 10.25 \pm 0.40         $ \\
NGC1332     & $0.68 \pm 0.12$~\cite{vandenBosch:2016}  & $ 22.30 \pm  1.85 $~\cite{Terrazas:2016b} & $9.17 ^{+0.06 }_{-0.06 }$~\cite{Kormendy:2013}   		&  $ 11.44 \pm	0.08  $  &  $8.80 \pm 0.13 $  &   $ 10.91 ^{+0.26}_{-0.35} $ \\	  
NGC1358     & $0.94 \pm 0.12$~\cite{vandenBosch:2016}  & $ 48.20 \pm  4.80 $~\cite{vandenBosch:2016} 	& $8.37 ^{+0.32 }_{-0.32 }$~\cite{vandenBosch:2016}   	&  $ 11.28 \pm	0.11  $  &  $9.00 \pm 0.09 $  &   $ 		  ...               $ \\
NGC1398     & $0.91 \pm 0.11$~\cite{vandenBosch:2016}  & $ 24.77 \pm  4.13 $~\cite{Terrazas:2016b} & $8.03 ^{+0.08 }_{-0.08 }$~\cite{Terrazas:2016b}  &  $ 11.73 \pm	0.15  $  &  $9.71 \pm 0.15 $  &   $ 10.57 \pm	0.20        $ \\ 
NGC1497     & $0.73 \pm 0.05$~\cite{vandenBosch:2016}  & $ 75.30 \pm  7.50 $~\cite{vandenBosch:2016}	& $8.63 ^{+0.19 }_{-0.19 }$~\cite{vandenBosch:2016}   	&  $ 11.40 \pm	0.08  $  &  $8.55 \pm 0.15 $  &   $ 		  ...               $ \\
NGC1667     & $0.69 \pm 0.08$~\cite{vandenBosch:2016}  & $ 56.10 \pm  5.60 $~\cite{vandenBosch:2016}	& $8.20 ^{+0.23 }_{-0.23 }$~\cite{vandenBosch:2016}   	&  $ 11.33 \pm	0.10  $  &  $9.51 \pm 0.15 $  &   $ 		  ...               $ \\
NGC1961     & $1.20 \pm 0.07$~\cite{vandenBosch:2016}  & $ 48.60 \pm  4.90 $~\cite{vandenBosch:2016}	& $8.29 ^{+0.34 }_{-0.34 }$~\cite{vandenBosch:2016}   	&  $ 11.79 \pm	0.10  $  &  $10.42\pm 0.10 $  &   $ 		  ...               $ \\
NGC2179     & $0.53 \pm 0.08$~\cite{vandenBosch:2016}  & $ 35.80 \pm  3.60 $~\cite{vandenBosch:2016}	& $8.31 ^{+0.23 }_{-0.23 }$~\cite{vandenBosch:2016}   	&  $ 10.85 \pm	0.10  $  &  $8.77 \pm 0.14 $  &   $ 		  ...               $ \\
NGC2273     & $0.60 \pm 0.09$~\cite{vandenBosch:2016}  & $ 29.50 \pm  1.90 $~\cite{Terrazas:2016b} & $6.94 ^{+0.02 }_{-0.02 }$~\cite{Kormendy:2013}   		&  $ 10.95 \pm	0.07  $  &  $9.24 \pm 0.07 $  &   $ 9.98  \pm	0.20        $ \\ 
NGC2787     & $-0.02\pm 0.07$~\cite{vandenBosch:2016}  & $ 7.45  \pm  1.24 $~\cite{Terrazas:2016b} & $7.61 ^{+0.04 }_{-0.06 }$~\cite{Kormendy:2013}   		&  $ 10.24 \pm	0.13  $  &  $8.43 \pm 0.18 $  &   $ 9.13  \pm	0.26        $ \\ 
NGC2911     & $0.88 \pm 0.12$~\cite{vandenBosch:2016}  & $ 43.50 \pm  4.30 $~\cite{vandenBosch:2016} 	& $9.09 ^{+0.29 }_{-0.29 }$~\cite{vandenBosch:2016}   	&  $ 11.32 \pm	0.10  $  &  $9.04 \pm 0.11 $  &   $ 		  ...               $ \\
NGC2960     & $0.73 \pm 0.08$~\cite{vandenBosch:2016}  & $ 67.10 \pm  7.12 $~\cite{Terrazas:2016b} & $7.03 ^{+0.02 }_{-0.02 }$~\cite{Kormendy:2013}   		&  $ 11.12 \pm	0.09  $  &  $9.31 \pm 0.12 $  &   $ 10.44 \pm	0.36        $ \\ 
NGC2974     & $0.55 \pm 0.07$~\cite{vandenBosch:2016}  & $ 21.50 \pm  2.40 $~\cite{vandenBosch:2016} 	& $8.23 ^{+0.09 }_{-0.09 }$~\cite{vandenBosch:2016}   	&  $ 11.10 \pm	0.09  $  &  $8.72 \pm 0.14 $  &   $ 10.23 \pm	0.13        $ \\ 
NGC3079     & $0.63 \pm 0.05$~\cite{vandenBosch:2016}  & $ 15.90 \pm  1.20 $~\cite{vandenBosch:2016} 	& $6.40 ^{+0.05 }_{-0.05 }$~\cite{vandenBosch:2016}   	&  $ 10.83 \pm	0.07  $  &  $9.95 \pm 0.07 $  &   $ 9.92  \pm	0.25        $ \\ 
NGC3081     & $0.66 \pm 0.13$~\cite{vandenBosch:2016}  & $ 33.50 \pm  3.40 $~\cite{vandenBosch:2016} 	& $7.20 ^{+0.30 }_{-0.30 }$~\cite{vandenBosch:2016}   	&  $ 10.90 \pm	0.11  $  &  $9.41 \pm 0.12 $  &   $ 		  ...               $ \\
NGC3115     & $0.42 \pm 0.06$~\cite{vandenBosch:2016}  & $ 9.54  \pm  0.40 $~\cite{Terrazas:2016b} & $8.95 ^{+0.03 }_{-0.13 }$~\cite{Kormendy:2013}   		&  $ 11.04 \pm	0.04  $  &  $9.02 \pm 0.11 $  &   $ 10.19 \pm 0.15         $ \\
NGC3227     & $1.03 \pm 0.14$~\cite{vandenBosch:2016}  & $ 23.75 \pm  2.63 $~\cite{Terrazas:2016b} & $7.32 ^{+0.14 }_{-0.23 }$~\cite{Kormendy:2013}   		&  $ 11.16 \pm	0.13  $  &  $9.40 \pm 0.11 $  &   $ 10.04 \pm 0.17         $ \\
NGC3368     & $0.49 \pm 0.06$~\cite{vandenBosch:2016}  & $ 10.40 \pm  0.96 $~\cite{Terrazas:2016b} & $6.88 ^{+0.09 }_{-0.09 }$~\cite{Kormendy:2013}   		&  $ 10.89 \pm	0.08  $  &  $9.18 \pm 0.09 $  &   $ 9.81  \pm 0.10         $ \\
NGC3379     & $0.48 \pm 0.09$~\cite{vandenBosch:2016}  & $ 10.70 \pm  0.54 $~\cite{Terrazas:2016b} & $8.62 ^{+0.11 }_{-0.11 }$~\cite{Kormendy:2013}   		&  $ 11.03 \pm	0.06  $  &  $7.28 \pm 0.12 $  &   $ 10.67 \pm 0.15         $ \\
NGC3393     & $0.78 \pm 0.11$~\cite{vandenBosch:2016}  & $ 49.20 \pm  8.19 $~\cite{Terrazas:2016b} & $7.20 ^{+0.27 }_{-0.27 }$~\cite{Kormendy:2013}   		&  $ 11.13 \pm	0.15  $  &  $9.83 \pm 0.15 $  &   $ 10.23 \pm 0.12         $ \\
NGC3414     & $0.46 \pm 0.08$~\cite{vandenBosch:2016}  & $ 25.20 \pm  2.70 $~\cite{vandenBosch:2016} 	& $8.40 ^{+0.07 }_{-0.07 }$~\cite{vandenBosch:2016}   	&  $ 11.02 \pm	0.11  $  &  $8.35 \pm 0.12 $  &   $ 10.47 \pm 0.15         $ \\
NGC3489     & $0.18 \pm 0.08$~\cite{vandenBosch:2016}  & $ 12.10 \pm  0.84 $~\cite{Terrazas:2016b} & $6.77 ^{+0.06 }_{-0.06 }$~\cite{Kormendy:2013}   		&  $ 10.50 \pm	0.07  $  &  $7.43 \pm 0.09 $  &   $ 9.62 ^{+0.23}_{-0.26}  $ \\
NGC3504     & $0.58 \pm 0.04$ 						   & $ 32.40 \pm  2.10 $~\cite{Nguyen:2020} & $7.20 ^{+0.16 }_{-0.11 }$~\cite{Nguyen:2020}  &  $ 11.06 \pm	0.06  $  &  $9.10 \pm 0.08 $  &   $ 		  ...               $ \\
NGC3801     & $0.78 \pm 0.09$~\cite{vandenBosch:2016}  & $ 46.30 \pm  4.60 $~\cite{vandenBosch:2016} 	& $8.28 ^{+0.31 }_{-0.31 }$~\cite{vandenBosch:2016}   	&  $ 11.26 \pm	0.09  $  &  $9.25 \pm 0.11 $  &   $ 		  ...               $ \\
NGC3992     & $1.07 \pm 0.13$~\cite{vandenBosch:2016}  & $ 15.30 \pm  1.50 $~\cite{vandenBosch:2016} 	& $7.51 ^{+0.28 }_{-0.28 }$~\cite{vandenBosch:2016}   	&  $ 11.09 \pm	0.12  $  &  $9.48 \pm 0.10 $  &   $ 		  ...               $ \\
NGC3998     & $0.11 \pm 0.06$~\cite{vandenBosch:2016}  & $ 14.30 \pm  1.25 $~\cite{Terrazas:2016b} & $8.93 ^{+0.04 }_{-0.03 }$~\cite{Kormendy:2013}   		&  $ 10.77 \pm	0.07  $  &  $8.53 \pm 0.13 $  &   $ 10.66 \pm	0.15        $ \\ 
NGC4151     & $0.60 \pm 0.12$~\cite{vandenBosch:2016}  & $ 20.00 \pm  2.80 $~\cite{vandenBosch:2016} 	& $7.81 ^{+0.08 }_{-0.08 }$~\cite{vandenBosch:2016}   	&  $ 10.85 \pm	0.13  $  &  $9.52 \pm 0.13 $  &   $ 10.27 \pm	0.15        $ \\ 
NGC4203     & $0.42 \pm 0.13$~\cite{vandenBosch:2016}  & $ 14.10 \pm  1.40 $~\cite{vandenBosch:2016} 	& $7.82 ^{+0.26 }_{-0.26 }$~\cite{vandenBosch:2016}   	&  $ 10.71 \pm	0.11  $  &  $9.34 \pm 0.10 $  &   $ 		  ...               $ \\
NGC4258     & $0.64 \pm 0.08$~\cite{vandenBosch:2016}  & $ 7.27  \pm  0.50 $~\cite{Terrazas:2016b} & $7.577^{+0.005}_{-0.005}$~\cite{Kormendy:2013}   		&  $ 10.83 \pm	0.08  $  &  $9.57 \pm 0.08 $  &   $ 10.05 \pm	0.18        $ \\ 
NGC4303     & $0.81 \pm 0.11$~\cite{vandenBosch:2016}  & $ 17.90 \pm  1.80 $~\cite{vandenBosch:2016} 	& $6.51 ^{+0.74 }_{-0.74 }$~\cite{vandenBosch:2016}   	&  $ 11.03 \pm	0.12  $  &  $9.72 \pm 0.09 $  &   $ 9.42  \pm	0.10        $ \\ 
NGC4388     & $0.75 \pm 0.08$~\cite{vandenBosch:2016}  & $ 16.53 \pm  1.60 $~\cite{Terrazas:2016b} & $6.86 ^{+0.01 }_{-0.01 }$~\cite{Kormendy:2013}   		&  $ 10.63 \pm	0.08  $  &  $8.96 \pm 0.10 $  &   $ 10.07 \pm	0.22        $ \\ 
NGC4472     & $0.89 \pm 0.11$~\cite{vandenBosch:2016}  & $ 17.14 \pm  0.59 $~\cite{Terrazas:2016b} & $9.40 ^{+0.10 }_{-0.02 }$~\cite{Kormendy:2013}   		&  $ 11.83 \pm	0.07  $  &  $9.25 \pm 0.11 $  &   $ 11.59 ^{+0.04}_{-0.07} $ \\ 
NGC4501     & $0.72 \pm 0.04$~\cite{vandenBosch:2016}  & $ 16.50 \pm  1.14 $~\cite{Terrazas:2016b} & $7.30 ^{+0.08 }_{-0.08 }$~\cite{Terrazas:2016b}  &  $ 11.26 \pm	0.07  $  &  $9.25 \pm 0.08 $  &   $ 10.11 \pm	0.16        $ \\ 
NGC4507     & $0.62 \pm 0.06$~\cite{vandenBosch:2016}  & $ 47.00 \pm  4.70 $~\cite{vandenBosch:2016} 	& $7.18 ^{+0.35 }_{-0.35 }$~\cite{vandenBosch:2016}   	&  $ 11.05 \pm	0.09  $  &  $9.64 \pm 0.13 $  &   $ 		  ...               $ \\
NGC4594     & $0.74 \pm 0.08$~\cite{vandenBosch:2016}  & $ 9.87  \pm  0.82 $~\cite{Terrazas:2016b} & $8.82 ^{+0.03 }_{-0.03 }$~\cite{Kormendy:2013}   		&  $ 11.46 \pm	0.08  $  &  $8.39 \pm 0.10 $  &   $ 10.81 \pm	0.20        $ \\ 
NGC4636     & $0.96 \pm 0.08$~\cite{vandenBosch:2016}  & $ 13.70 \pm  1.40 $~\cite{vandenBosch:2016} 	& $8.58 ^{+0.22 }_{-0.22 }$~\cite{vandenBosch:2016}   	&  $ 11.27 \pm	0.09  $  &  $8.49 \pm 0.14 $  &   $ 		  ...               $ \\
NGC4699     & $0.49 \pm 0.12$~\cite{vandenBosch:2016}  & $ 18.90 \pm  2.05 $~\cite{Terrazas:2016b} & $8.25 ^{+0.05 }_{-0.05 }$~\cite{Terrazas:2016b}  &  $ 11.40 \pm	0.11  $  &  $9.35 \pm 0.10 $  &   $ 11.12 \pm 0.26         $ \\
NGC4736     & $0.00 \pm 0.12$~\cite{vandenBosch:2016}  & $ 5.00  \pm  0.79 $~\cite{Terrazas:2016b} & $6.83 ^{+0.10 }_{-0.10 }$~\cite{Kormendy:2013}   		&  $ 10.59 \pm	0.16  $  &  $8.38 \pm 0.15 $  &   $ 9.89  \pm 0.09         $ \\
NGC4826     & $0.38 \pm 0.12$~\cite{vandenBosch:2016}  & $ 7.27  \pm  1.18 $~\cite{Terrazas:2016b} & $6.19 ^{+0.11 }_{-0.11 }$~\cite{Kormendy:2013}   		&  $ 10.60 \pm	0.19  $  &  $8.71 \pm 0.15 $  &   $ 9.55  \pm 0.22         $ \\
NGC4945     & $0.77 \pm 0.12$~\cite{vandenBosch:2016}  & $ 3.58            $~\cite{Kormendy:2013} 		& $6.13 ^{+0.22 }_{-0.15 }$~\cite{Kormendy:2013}   		&  $ 10.76 \pm	0.12  $  &  $9.28 \pm 0.07 $  &   $ 9.39  \pm 0.19         $ \\
NGC5005     & $0.54 \pm 0.06$~\cite{vandenBosch:2016}  & $ 14.60 \pm  1.50 $~\cite{vandenBosch:2016} 	& $8.27 ^{+0.23 }_{-0.23 }$~\cite{vandenBosch:2016}   	&  $ 11.14 \pm	0.09  $  &  $8.88 \pm 0.11 $  &   $ 		  ...               $ \\
NGC5018     & $0.62 \pm 0.06$~\cite{vandenBosch:2016}  & $ 40.55 \pm  4.87 $~\cite{Terrazas:2016b} & $8.02 ^{+0.08 }_{-0.08 }$~\cite{Terrazas:2016b}  &  $ 11.58 \pm	0.09  $  &  $9.19 \pm 0.19 $  &   $ 10.98 \pm	0.27        $ \\ 
NGC5055     & $0.77 \pm 0.09$~\cite{vandenBosch:2016}  & $ 8.70  \pm  0.90 $~\cite{vandenBosch:2016} 	& $8.92 ^{+0.10 }_{-0.10 }$~\cite{vandenBosch:2016}  	&  $ 11.03 \pm	0.10  $  &  $9.59 \pm 0.10 $  &   $ 10.49 \pm	0.11        $ \\ 
NGC5127     & $0.96 \pm 0.07$~\cite{vandenBosch:2016}  & $ 62.50 \pm  6.30 $~\cite{vandenBosch:2016} 	& $8.27 ^{+0.41 }_{-0.41 }$~\cite{vandenBosch:2016}  	&  $ 11.38 \pm	0.10  $  &  $9.61 \pm 0.15 $  &   $ 		  ...               $ \\
NGC5248     & $0.69 \pm 0.06$~\cite{vandenBosch:2016}  & $ 17.90 \pm  1.80 $~\cite{vandenBosch:2016} 	& $6.30 ^{+0.38 }_{-0.38 }$~\cite{vandenBosch:2016}  	&  $ 10.92 \pm	0.09  $  &  $9.67 \pm 0.10 $  &   $ 		  ...               $ \\
NGC5252     & $0.88 \pm 0.06$~\cite{vandenBosch:2016}  & $ 103.70\pm  10.40$~\cite{vandenBosch:2016} 	& $9.07 ^{+0.34 }_{-0.34 }$~\cite{vandenBosch:2016}  	&  $ 11.52 \pm	0.09  $  &  $9.34 \pm 0.12 $  &   $ 10.85 \pm	0.26        $ \\ 
NGC5495     & $1.11 \pm 0.15$~\cite{vandenBosch:2016}  & $ 126.30\pm  11.60$~\cite{vandenBosch:2016} 	& $7.00 ^{+0.05 }_{-0.05 }$~\cite{vandenBosch:2016}  	&  $ 11.76 \pm	0.14  $  &  $9.81 \pm 0.09 $  &   $ 10.54 \pm	0.12        $ \\ 
NGC5695     & $0.64 \pm 0.08$~\cite{vandenBosch:2016}  & $ 54.60 \pm  5.50 $~\cite{vandenBosch:2016} 	& $8.00 ^{+0.32 }_{-0.32 }$~\cite{vandenBosch:2016}  	&  $ 10.94 \pm	0.10  $  &  $8.79 \pm 0.14 $  &   $ 		 ...                $ \\
NGC5728     & $0.90 \pm 0.08$~\cite{vandenBosch:2016}  & $ 37.60 \pm  3.80 $~\cite{vandenBosch:2016} 	& $8.05 ^{+0.29 }_{-0.29 }$~\cite{vandenBosch:2016}  	&  $ 11.31 \pm	0.11  $  &  $9.46 \pm 0.12 $  &   $ 		 ...                $ \\
NGC0584     & $0.49 \pm 0.10$~\cite{vandenBosch:2016}  & $ 19.10 \pm  1.00 $~\cite{Thater:2019} & $8.13 ^{+0.16 }_{-0.16 }$~\cite{Thater:2019}  &  $ 11.02 \pm	0.10  $  &  $8.17 \pm 0.16 $  &   $ 		 ...                $ \\
NGC5846     & $0.80 \pm 0.10$~\cite{vandenBosch:2016}  & $ 24.90 \pm  2.30 $~\cite{vandenBosch:2016} 	& $9.04 ^{+0.06 }_{-0.06 }$~\cite{vandenBosch:2016}   	&  $ 11.52 \pm	0.09  $  &  $8.54 \pm 0.13 $  &   $ 11.10 \pm	0.15        $ \\ 
NGC6240(S)  & $0.93 \pm 0.08$~\cite{vandenBosch:2016}  & $ 105.00\pm  10.50$~\cite{vandenBosch:2016} 	& $9.17 ^{+0.21 }_{-0.21 }$~\cite{vandenBosch:2016}   	&  $ 11.94 \pm	0.10  $  &  $9.92 \pm 0.17 $  &   $ 		 ...                $ \\
NGC7331     & $0.57 \pm 0.08$~\cite{vandenBosch:2016}  & $ 12.20 \pm  1.20 $~\cite{vandenBosch:2016} 	& $8.02 ^{+0.18 }_{-0.18 }$~\cite{vandenBosch:2016}   	&  $ 11.07 \pm	0.10  $  &  $9.69 \pm 0.09 $  &   $ 		 ...                $ \\
NGC7332     & $0.40 \pm 0.07$~\cite{vandenBosch:2016}  & $ 21.70 \pm  2.20 $~\cite{vandenBosch:2016} 	& $7.08 ^{+0.18 }_{-0.18 }$~\cite{vandenBosch:2016}   	&  $ 10.76 \pm	0.08  $  &  $8.33 \pm 0.19 $  &   $ 10.22 \pm	0.34        $ \\ 
NGC7582     & $0.89 \pm 0.19$~\cite{vandenBosch:2016}  & $ 22.30 \pm  9.85 $~\cite{Terrazas:2016b} & $7.74 ^{+0.10 }_{-0.07 }$~\cite{Kormendy:2013}   		&  $ 11.20 \pm	0.32  $  &  $9.55 \pm 0.40 $  &   $ 10.15 \pm	0.20        $ \\ 
UGC3789     & $0.66 \pm 0.11$~\cite{vandenBosch:2016}  & $ 49.90 \pm  5.42 $~\cite{Terrazas:2016b} & $6.98 ^{+0.07 }_{-0.07 }$~\cite{Kormendy:2013}   		&  $ 10.92 \pm	0.12  $  &  $9.00 \pm 0.10 $  &   $ 10.18 \pm	0.14        $ \\ 
\hline 
\end{longtable}
\vspace{-1cm}
\begin{tablenotes}
\footnotesize               
\item[*]  Log $M_{\rm HI}$ is derived from corrected 21-cm line flux ($m21c$) collected from  \textit{HyperLeda}~\cite{Makarov:2014}
\item[**] Log $M_{\rm bulge}$ is collected from Ref~\cite{Bohn:2020}.
\end{tablenotes}
\end{ThreePartTable}
\end{small}

\vspace{0.3cm}

\begin{small}
\setlength{\LTcapwidth}{52in}
\begin{ThreePartTable}
\setlength{\tabcolsep}{18pt}
\begin{longtable}{cccc}
\caption{\textbf{Best-fitted linear relations between $\mu_{\rm HI}$ and $M_{\rm BH}$ for different samples.}}
\label{etb2}\\
\vspace{-0.5cm}
\tabcolsep=0.9cm \\
\hline \hline 
    & HI-detected Galaxy sample & All Galaxy sample &  The BH sample\\ 
\midrule
\endfirsthead
$\alpha$ &$\rm {{-0.43\pm0.02}}$&$\rm {{-0.59\pm0.19}}$&$\rm {{-0.37\pm0.06}}$\\
$\beta$ & $\rm{{2.29\pm0.18}}$& $\rm{{3.20\pm1.38}}$&$\rm{{1.30\pm0.50}}$\\
\hline 
\end{longtable}
\vspace{-1cm}
\begin{tablenotes}
\item log $\mu_{\rm HI}$ =  $\alpha$ log$M_{\rm BH}$ + $\beta$               
\end{tablenotes}
\end{ThreePartTable}
\end{small}

\vspace{0.3cm}
\begin{small}
\setlength{\LTcapwidth}{6.2in}
\begin{ThreePartTable}
\setlength{\tabcolsep}{10pt}
\begin{longtable}{ccccccc}
\caption{\textbf{Partial correlation coefficients and variance contributions regarding the correlation between $\mu_{\rm HI}$ and major galactic parameters.}}
\label{etb3}\\
\vspace{-0.5cm}
\tabcolsep=0.4cm \\
\hline \hline 
    & $M_{\rm BH}$ & $M_{\star}$ & $\Sigma_{\rm star}$ & $M_{\rm bulge}$&SSFR&
    Variance(\%)\\ 
\midrule
\endfirsthead
$M_{\rm BH}$ &-/- & 
{{-0.35/-0.51}}&
{{-0.43/-0.64}}&
{{-0.17/-0.40}}&
{{-0.08/-0.54}}&
{{66.8/52.3}}\\
$M_{\star}$&
{{-0.13/-0.10}}&-/- &
{{-0.43/-0.58}}&
{{-0.10/-0.09}}&
{{-0.49/-0.38}}&
{{6.8/8.2}}\\
$\Sigma_{\rm star}$ &
{{-0.47/-0.20}} &
{{-0.55/-0.44}}  &-/-&
{{-0.61/-0.26}} &
{{-0.32/-0.44}}&
{{10.8/11.2}}\\
$M_{\rm bulge}$ & 
{{-0.18/-0.24}} &
{{-0.25/-0.45}}  &
{{-0.32/-0.52}} &-/-&
{{-0.27/-0.45}}&
{{15.6/28.2}}\\
SSFR&
{{0.45/0.17}}&
{{0.58/0.36}}&
{{0.43/0.57}}&
{{0.57/0.26}}&-/-&-/-\\
\hline 
\end{longtable}
\begin{tablenotes}
\footnotesize 
\vspace{-1cm}
 \item[]The second to fifth columns show the partial correlation coefficients between $f_{\rm HI}$ and each parameter in the first column by controlling the parameters in the first row. The last column shows the variance contribution to $f_{\rm HI}$ in each parameter from PLS Regression between $\mu_{\rm HI}$, $M_{\rm BH}$, $M_{\star}$, $\Sigma_{\rm star}$ and $M_{\rm bulge}$. The first and second values are for the BH and galaxy sample, respectively.        
\end{tablenotes}
\end{ThreePartTable}
\end{small}

\renewcommand{\arraystretch}{1.0}
\renewcommand{\baselinestretch}{1.5}

\clearpage
\setcounter{figure}{0}
\begin{Extended Data Figure}
	\centering
	\includegraphics[width=0.9\textwidth]{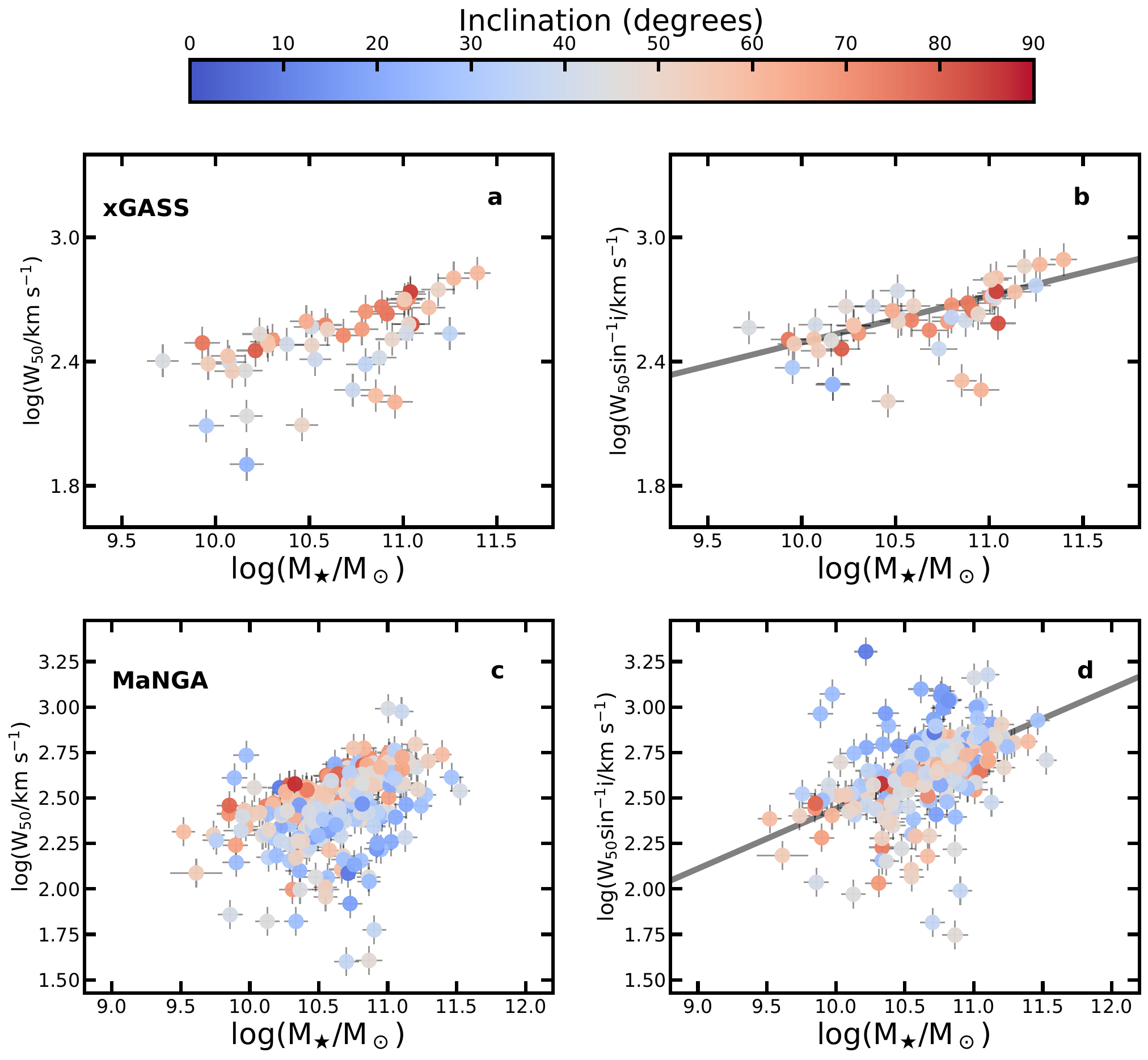} 
	\caption{\small{\textbf{The impact of inclinations on the observed HI line width.} 
The relation between $M_{\star}$ and the observed HI line width are shown in panels ({\bf a, c}), while the relation between $M_{\star}$ and the inclination-corrected HI line width are shown in panels({\bf b, d}), for MaNGA and xGASS galaxies respectively. The error bars refer to 1-$\sigma$ errors for log$M_{\star}$. For $W_{50}$, we denote 20\% measurement uncertainties.}}
	\label{edf1}
\end{Extended Data Figure}
\begin{Extended Data Figure}
	\centering
	\includegraphics[scale=0.5]{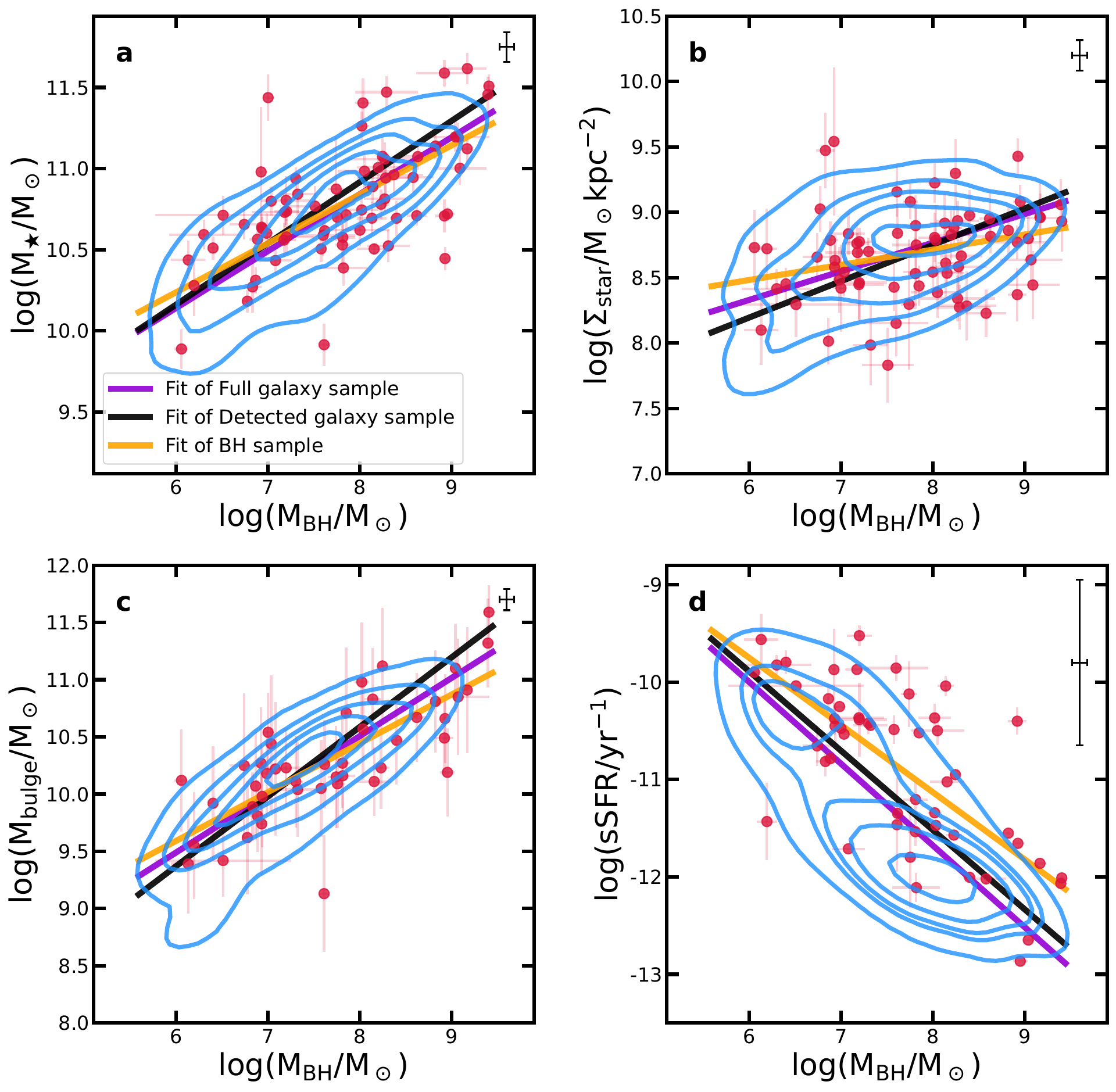} 
	\caption{\small{\textbf{The relation between BH masses and other major galactic parameters for the BH and galaxy sample.} The relation between $M_{\rm BH}$ and $M_{\star}$, $\Sigma_{\rm star}$, $M_{\rm bulge}$, and $\rm SSFR$  for the BH sample (red filled dots) and the full galaxy sample (blue contours) are shown in Panels {\bf a}, {\bf b}, {\bf c}, and {\bf d}, respectively. The best-fitted relation for the galaxy sample, the full galaxy, and the HI-detected galaxy sample are respectively drawn in solid orange, darkviolet, and black lines in each panel. These relations are used to derive the residuals of the corresponding galactic parameters after controlling for $M_{\rm BH}$ in the bottom row of Figure~\ref{fig3}, and Extended Data Figure~\ref{edf3}. The median error of the galaxy sample is shown in the upper right in each panel. The error bars refer to 1-$\sigma$ errors.}}
	\label{edf2}
\end{Extended Data Figure}

\begin{Extended Data Figure}
	\centering
	\includegraphics[scale=0.20]{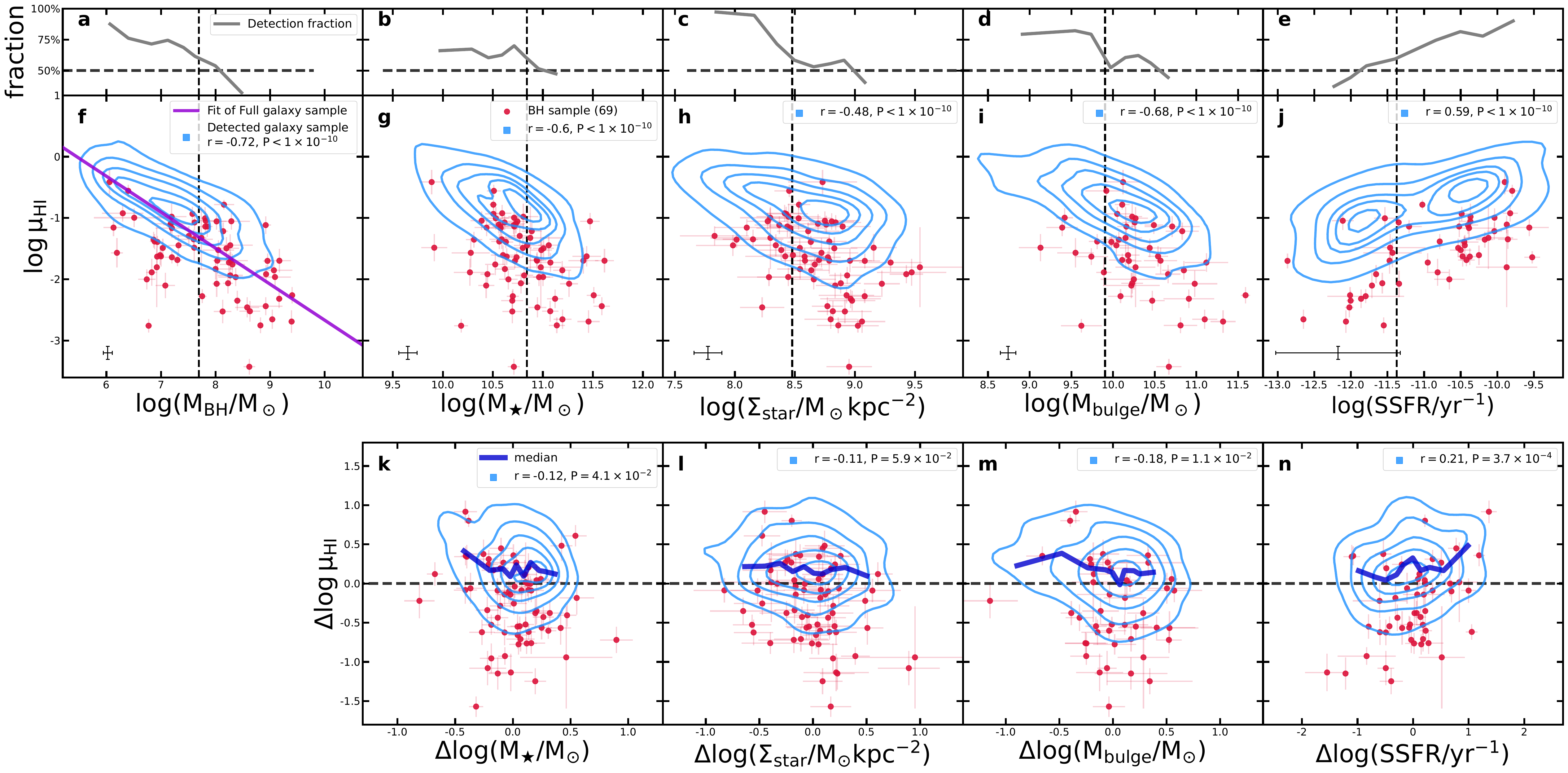} 
	\caption{\small{\textbf{The impact of $M_{\rm BH}$ on the correlation between $\mu_{\rm HI}$ and other major galactic parameters for the full galaxy sample.} Similar as Figure~\ref{fig3}, but with the third row replaced by the relation between the residual in $\mu_{\rm HI}$ and other galactic parameters after removing their dependence on $M_{\rm BH}$ based on the full sample instead of only the HI-detected ones. The solid medium-blue lines denote the running median of the residuals in $\mu_{\rm HI}$. Since the HI-detected samples are always biased compared to the full sample, the running medians exhibits a positive offset compared to zero values(black dashed lines). The median error of the galaxy sample is shown in the lower left in panel ({\bf f $\sim$ j}). The error bars refer to 1-$\sigma$ errors. In the top-right corners of the middle and bottom panels, we show the Spearman correlation coefficients for the HI-detected galaxy sample between the corresponding $x$ and $y$ variables.}}
	\label{edf3}
\end{Extended Data Figure}

\begin{Extended Data Figure}
	\centering
	\includegraphics[scale=0.45]{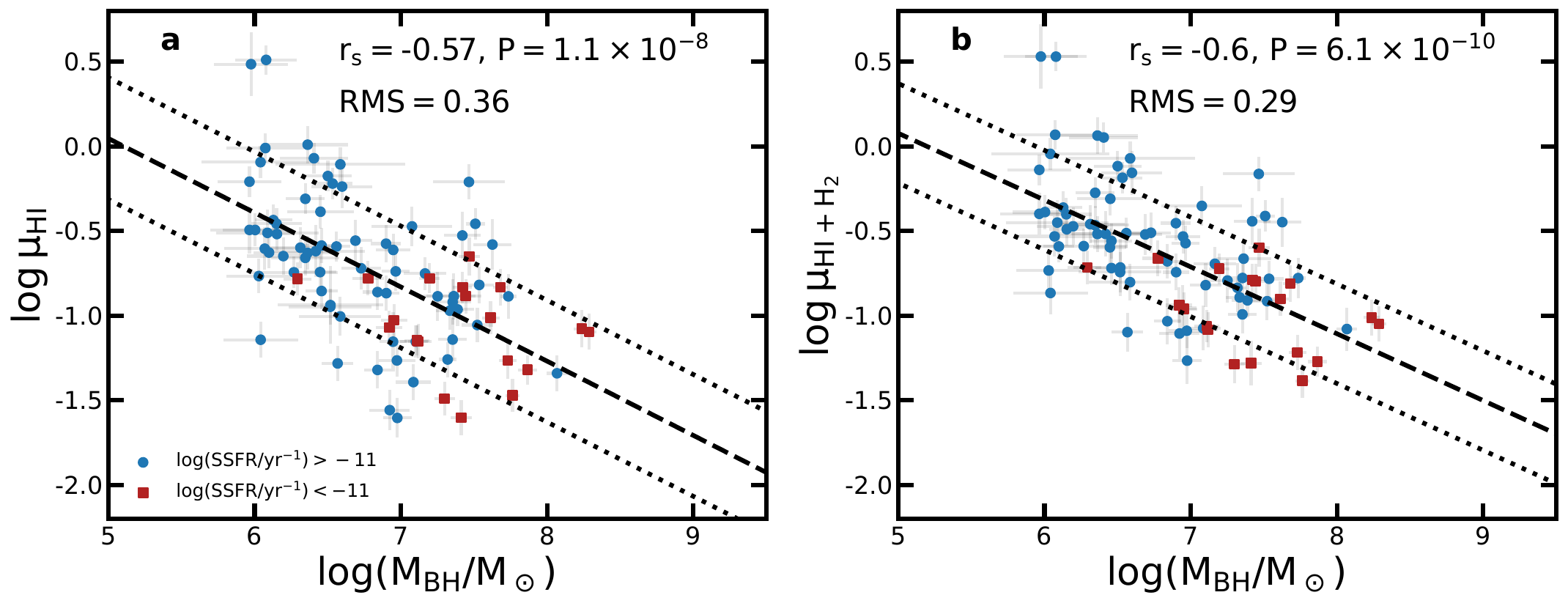} 
	\caption{\small{\textbf{Comparison between the relations of BH masses to the HI gas fraction and to the total gas fraction.} 
The relation between $M_{\rm BH}$ and HI gas fraction is shown in panel {\bf a}, and the relation between  $M_{\rm BH}$ and total gas fraction is shown in panel {\bf b} for central galaxies. The blue circles and red squares denote respectively star-forming and quiescent galaxies, which are classified based on their SSFR. Spearman correlation coefficients and the RMS between data and the fitted relations (black dashed lines) are shown in the top-right corners. The error bars refer to 1-$\sigma$ errors.}}
	\label{edf4}
\end{Extended Data Figure}

\begin{Extended Data Figure}
	\centering
	\includegraphics[scale=0.45]{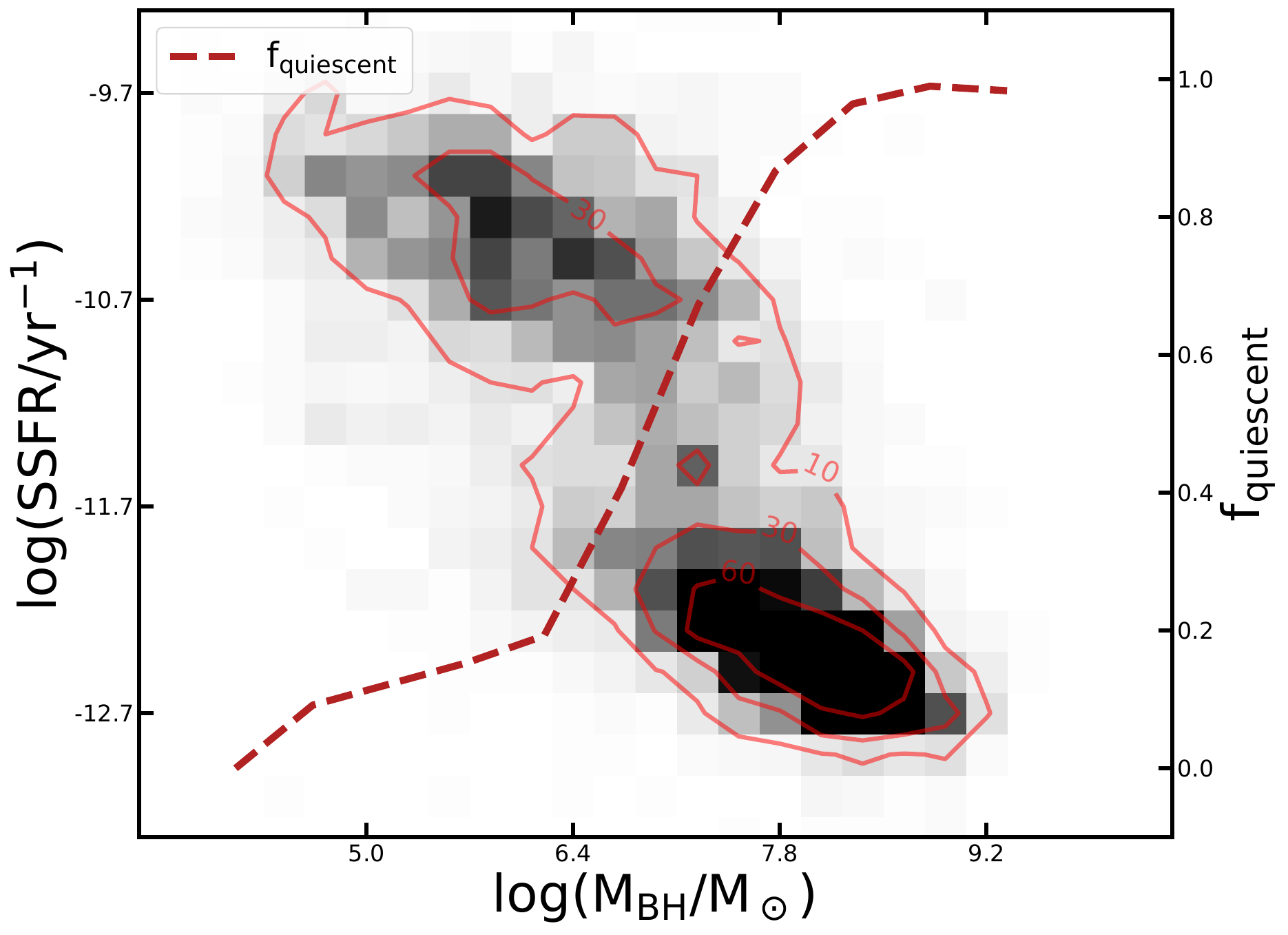} 
	\caption{\small{\textbf{Relation between SSFR and BH masses for SDSS group central galaxies.} Density map of SDSS group central galaxies is shown in the SSFR-$M_{\rm BH}$ plot. The dashed line denotes the quiescent fraction as a function of $M_{\rm BH}$. The contours and corresponding numbers denote the galaxy number for each pixel.}}
	\label{edf5}
\end{Extended Data Figure}

\begin{Extended Data Figure}
	\centering
	\includegraphics[scale=0.8]{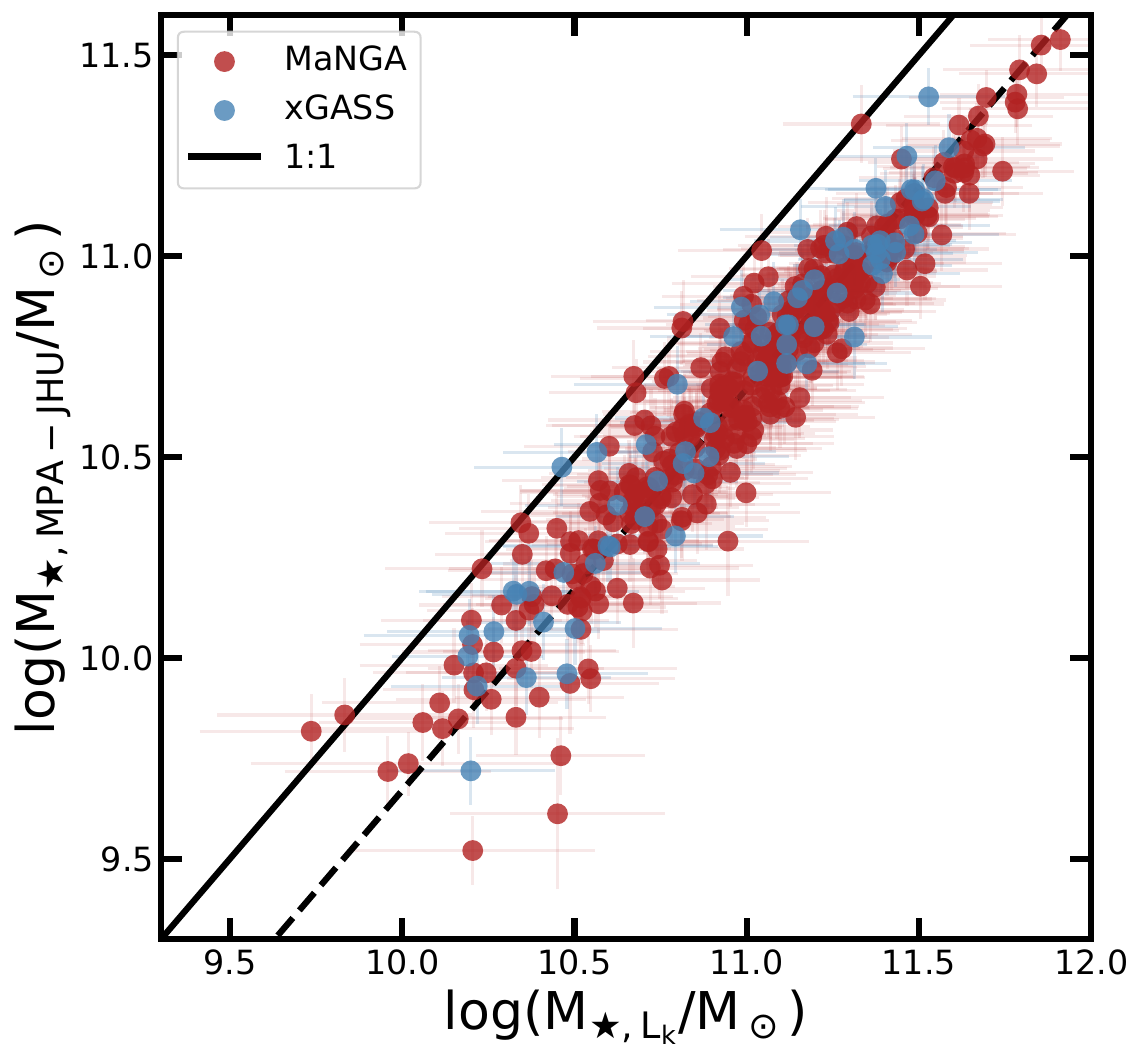} 
	\caption{\small{\textbf{The comparison between the stellar masses derived from SED-fitting and those derived based on K-band luminosities}. 
The SED-based estimation $M_{\star}$ is taken from the MPA-JHU catalog and the K-band luminosity based estimation of $M_{\star}$ is derived from the relation among $M_\star$, $L_{\rm K}$ and $R_{\rm e}$ given by ref.~\cite{vandenBosch:2016}. The dashed line denotes the `1:1' line shifted by the mean mass difference of 0.32\,dex. The error bars refer to 1-$\sigma$ error.}}
	\label{edf6}
\end{Extended Data Figure}

\end{bibunit}

\end{document}